\documentclass{article}

\usepackage[utf8]{inputenc}
\usepackage[square,numbers]{natbib}
\bibpunct{(}{)}{}{a}{}{;}
\usepackage{hyperref}
\usepackage{tikz}
\usepackage{subcaption}
\usepackage{hyperref}
\usepackage{adjustbox}
\usepackage[labelformat=parens,labelsep=quad, skip=3pt]{caption}
\usepackage{tabulary}
\usepackage{floatrow}
\usepackage{authblk}
\usepackage{siunitx}
\usepackage{soul}
\usepackage{amsmath}
\usepackage{amsfonts}
\usepackage{multirow}
\usepackage{booktabs}
\usepackage{fancyhdr}
\usepackage{algorithm,xcolor,caption}

\DeclareCaptionFont{lightblue}{\color{blue}}
\usepackage{bbm}
\usepackage[colorinlistoftodos]{todonotes}
\newfloatcommand{capbtabbox}{table}[][\FBwidth]

\usepackage[margin=0.25cm]{caption}
\usepackage{blindtext}

\graphicspath{ {./images/} }
\usepackage{todonotes}
\usepackage{tikz}

\defcitealias{UNISDR}{UNISDR, 2009}
\defcitealias{Report_2017}{Presidenza del Consiglio dei Ministri}

\newcommand{\mycitet}[1]{\citetalias{#1} (\citeyear{#1})}
\newcommand{\mycitep}[1]{\citepalias[\citeyear{#1}]{#1}}

\usepackage{geometry}
\title{Social and material vulnerability in the face of seismic hazard: an analysis of the Italian case}

\author{Oleksandr Didkovskyi\(^{\dag, \sharp}\), Giovanni Azzone\(^{*,\;\sharp}\), \newline \medskip Alessandra Menafoglio\(^\dag\),  Piercesare Secchi\(^{\dag,\;\sharp}\)\newline 

\(^\dag\) MOX - Dipartimento di Matematica, Politecnico di Milano, Italy\newline
\(^*\) Dipartimento di Ingegneria Gestionale, Politecnico di Milano, Italy\newline
\(^\sharp\) Center for Analysis, Decisions and Society, Human Technopole, Italy}

\date{}
\begin{document}
\maketitle
\begin{abstract}
The assessment of the vulnerability of a community endangered by seismic hazard is of paramount importance for planning a precision policy aimed at the prevention and reduction of its seismic risk. We aim at measuring the vulnerability of the Italian municipalities exposed to seismic hazard, by analyzing the open data offered by the Mappa dei Rischi dei Comuni Italiani provided by ISTAT, the Italian National Institute of Statistics. Encompassing the Index of Social and Material Vulnerability already computed by ISTAT, we also consider as referents of the latent social and material vulnerability of a community, its demographic dynamics and the age of the building stock where the community resides. Fusing the analyses of different indicators, within the context of seismic risk we offer a tentative ranking  of the Italian municipalities in terms of their social and material vulnerability, together with differential profiles of their dominant fragilities which constitute the basis for planning precision policies aimed at seismic risk prevention and reduction.

\end{abstract}

\section{Introduction}
After the tragic sequence of earthquakes near Amatrice in 2016, the Italian government set up the Casa Italia task force\footnote{The task force Casa Italia of the Italian Presidency of the Council of Ministers was established on September 23, 2016. Its members were Giovanni Azzone (Project Manager and Scientific Director), Massimo Alvisi, Michela Arnaboldi, Alessandro Balducci, Marco Cammelli, Guido Corso, Francesco Curci, Daniela De Leo, Carlo Doglioni, Andrea Flori, Manuela Grecchi, Massimo Livi Bacci, Maurizio Milan, Alessandra Menafoglio, Pietro Petraroia, Fabio Pammolli, Davide Rampello, Piercesare Secchi. The 3rd of July 2017,  the Italian Presidency of the Council of Ministers estabilished Casa Italia as one of its departments, committed to the prevention against natural risks  (\url{http://www.casaitalia.governo.it/it/}). The task force Casa Italia finished its mission on May 31, 2018, while the department Casa Italia is still continuing its activities. The authors acknowledge the task force Casa Italia for the scientific discussions that were inspirational to the present work.}
to develop a plan for housing and land care directed to the better protection of citizens, and public and private goods, against natural risks.
One of the first activities of Casa Italia was aimed at integrating and enhancing the rich information on natural risks already available as the result of numerous and continuing investigations carried out by several national institutions\footnote{The institutions consulted by the task force Casa Italia were, in alphabetical order: CNR - Consiglio Nazionale delle Ricerche; ENEA - Agenzia nazionale per le nuove tecnologie, l'energia e lo sviluppo economico sostenibile; INGV - Istituto Nazionale di Geofisica e Vulcanologia; ISPRA - Istituto Superiore per la Protezione e la Ricerca Ambientale; ISTAT- Isitituto Nazionale di Statistica;  MIBACT -Ministero per i beni e le attività culturali e per il turismo.}. The action was devoted to identify the sources of information and the databases allowing for a unified and integrated vision of the natural risks insisting on the Italian territory, with particular reference to the three factors that compose risk, namely, hazard, vulnerability and exposure. Consistently with the mission of Casa Italia, the survey was limited to databases that (a) were elaborated by official and national research institutes, (b) covered the entire national territory, and (c) had a spatial resolution sufficient to allow identification and comparison of local specificities. Believing that it is the community, more than the single individual, which should be the collective principal agent of the prevention endeavor against natural risks,  the municipality ({\em Comune}, in Italian) was identified as the smallest spatial statistical unit for the actual analyses. The product of this action is the {\em Mappa dei Rischi dei Comuni Italiani}\footnote{ \url{http://www4.istat.it/it/mappa-rischi}} (MRCI), which was presented to the public on February 18, 2019 by the Department Casa Italia -- a meanwhile established permanent division of the Italian Government, derived from the positive experience of the Casa Italia task force which terminated its activities in May 2018.  The MRCI is a freely accessible web portal implemented by ISTAT, the Italian institute providing official statistics of the country. The portal supplies integrated information on different natural risks of Italian municipalities -- such as earthquake, flooding, landslide, volcano eruption -- in conjunction with socio-economic and demographic data. With the aim of creating a widespread awareness of the fragility of the Italian territory, the MRCI offers the possibility of viewing and downloading indicators, charts and maps, together with guided interactive features for data searching and filtering.

Amongst the indicators of natural risks available in the MRCI, the focus of this work is on those allowing to quantify the risk from earthquake. As a matter of fact, prevention of seismic risk can only act on the reduction of the vulnerability of the communities endangerd by it, when the option of reducing their exposition by moving them to a less hazardous territory is excluded \mycitep{Report_2017}. To develop precision policies for  seismic risk prevention, it is therefore of paramount importance to understand and quantify the distinctive vulnerability of each community facing a seismic hazard, interpreted as its inability to withstand the consequences of a catastrophic seismic event 
\citepalias{UNISDR}

An emerging body of literature has been devoted to the assessment of social vulnerability, the interest in this topic being consistently high since the 1960th \citep{Cutter2003}. Typical analyses developed in this context were carried out accounting for several factors related with, e.g., population' age, ethnicity, and socioeconomic status, all these aspects contributing directly or indirectly to the the social and economic vulnerability of the population being studied. The complexity of defining social vulnerability through numerical indicators is well-recognized in the literature; for instance, \citet{Cutter2003} present 42 variables which contribute to societal characterization when dealing with its vulnerability. In fact, multiple authors have worked on developing ways to build summaries out of large sets of numerical indicators, e.g., through principal components analysis or factor analysis (see, the review of \citet{Cutter2003} and referenced therein). As a result of this approach, the principal components (or factors) are associated with one feature such as age, personal wealth, occupation, ethnic group, and then analyzed separately (e.g.,\citet{Cutter2003}), or merged within a global Social Vulnerability Index (SVI) score (e.g., by using an additive model, \cite{Cutter2003, Frigerio16b}). 
Having defined one or more vulnerability scores, their variability can be then statistically analyzed for evidencing spatial or temporal trends. (e.g., \citet{Cutter2003,cutter2008temporal, andrew2008social}).

The complexity of elaborating indicators of social vulnerability aside, for the scope of this work it is relevant to mention that a number of authors precisely focused on the analysis of social vulnerability in the face of natural hazards \citep{ Birkman2006, Yoon2012, Fatemi2017}, with particular reference to climate change \citep{adger1999social,vincent2004creating}, flood hazard \citep{fekete2009validation, chakraborty2020place} and sea-level rise \citep{wu2002vulnerability, wood2010community}. As far as seismic risk is concerned, most vulnerability studies focused on the \emph{physical} vulnerability of the residential buildings where people live (see, e.g., \citet{dolce2003earthquake, vicente2011seismic, neves2012seismic, maio2015seismic}), rather than on the vulnerability of the communities themselves, which is the aim of this work instead. 

We also note that a large part of previous studies addressed the problem at the local scale (e.g., at the level of single buildings) \citep{lang2002seismic, dolce2003earthquake, maio2015seismic}, or at a county level \citep{adger1999social,wu2002vulnerability, fekete2009validation, wood2010community, chakraborty2020place}, whereas only a few were developed at a country scale (see 
% Vietnam, Canada, USA, Germany
\citet{nicholls1993synthesis, vincent2004creating, turvey2007vulnerability}).
In this regard, it is worth mentioning that defining social vulnerability is a country-specific process, as it strongly depends on the peculiar features of the societal system being studied. This implies that conclusions should not be blindly applied to countries other than the one for which the analysis was originally made \citep{holand2013replicating}. As a matter of fact, very few works focused on the case of Italy \citep{Frigerio16,Frigerio16b, Frigerio18}, although the country is amongst the most exposed to natural
risks \citep{Garschagen2016}. 
To our knowledge, the recent study presented in \citet{FrigerioVentura2016} is the only available work that investigates social vulnerability with respect to seismic hazard in Italy. To capture social vulnerability, these authors studied  
through factor analysis 
a number of socio-economic indicators made available by ISTAT, and related with age, employment, education, and anthropization.  
The identified factors were then used to build a SVI, which was then discretized in four classes of vulnerability (high, medium, low, very low).

As highlighted in \citet{Frigerio18}, social vulnerability should not be interpreted as a stand-alone concept, but rather as context-dependent and associated with the extreme event considered -- earthquakes in this work. Considering the community standing on a seismic landscape as made of humans interacting with structures and infrastructures, is thus key to provide sensible assessments of its vulnerability to seismic events and drive decision makers to policy planning. In this work we explore the MRCI with the goal of eliciting from this database the multidimensional aspects outlining the  vulnerability of Italian municipalities facing seismic hazard.  
We shall approach this multifaceted problem by exploring the (spatial) variability of its diverse dimensions, and merging them in an overall \emph{multi-aspect} ranking of the municipalities, usable to support decision makers in designing public policies.
Unlike previous works, to define and measure the vulnerability of a community endangered by seismic hazard, we believe it is of the greatest importance to include a factor capturing the vulnerability of the dwelling occupied by the community. Hence, beside the consideration of the socio-economic factors related to its poverty, employment and education, and those related to its demographic dynamics, we will also consider the age distribution of the building stock lived in by the community as a key indicator influencing both its social and material vulnerability, being the age of the building stock directly related to the safety and quality of the places where the community lives. To emphasize this expanded concept of vulnerability in the face of seismic hazard, we refer to it as the {\em social and material} vulnerability of a community 
(consistently with the expression used by ISTAT)
and we consider it to be the determinant on which precision policies can act upon for the reduction of seismic risk. 
In fact, while the \textit{physical} side of vulnerability influences the expected impact on buildings of an earthquake of a given magnitude, the {\em social and material}  vulnerability of a community affects its resilience, and capacity to invest resources to improve the safety of buildings and to react to an adverse event.

The remaining of the paper is organized as follows. In Section \ref{sect:seismic_landscape} we set out the Italian seismic landscape through a summary indicator 
(the peak ground acceleration, PGA)
which is apt to describe the seismic hazard faced by Italian municipalities. This is the landscape the social communities live upon, and on this landscape we project our assessments of social and material vulnerability. Section \ref{sect:IVSM} introduces an index of social and material vulnerability 
(\(IVSM\)),
which is already produced by ISTAT. We will identify the spatial hot spots of \(IVSM\)  
by using LISA maps \citep{Anselin95, Anselin06}, and segment them according to seismic hazard. Arguing that \(IVSM\) is not sufficient to characterize the social and material vulnerability of the Italian municipalities for the purpose of designing precision policies for seismic risk prevention, we will then analyze two other factors which we believe have an impact on a community's fragility when confronted with a seismic event: the demographic growth and age of the residents, and the age of the building stock they live in. 
In both cases, the indicators shall be first analysed separately in terms of their spatial variability and hot spots, and then related with the Italian seismic landscape.
These analyses will be carried out in Section \ref{sect:demographics} and Section \ref{sect:building_stock}, respectively. Ultimately, in Section \ref{sect:aggregation}, we will aggregate this analyses with the spatial analysis of \(IVSM\) to provide a tentative ranking of Italian municipalities in terms of their social and material vulnerability to seismic hazard. 
To this end, we shall rely on the Copeland method \citep{copeland}, which allows one to provide a global ranking from the aggregation of single indicators.
A final section with further supporting arguments and analyses, and a section with conclusions, will close the paper.

Data and code used to elaborate the results reported in this paper are freely available at the following location: \url{https://github.com/alexdidkovskyi/MRCI_paper}

\section{The seismic hazard landscape} \label{sect:seismic_landscape}
We cannot act on and change seismic hazard. This is a given landscape social communities must live upon; on this landscape we should evaluate their vulnerability.

The seismic hazard at a given point on the Earth surface can be measured in different ways. In engineering application, a commonly used quantity, often plotted in seismic hazard maps, is the Peak Ground Acceleration (PGA), indicated in the literature as  $a(g)$ \citep{pga_def}. This represents the maximum horizontal acceleration on rigid soil which is exceeded with a given probability $p$, say $p=0.1$, in 50 years. The unit of measure of \(a(g)\) is \(g,\) the acceleration due to Earth's gravity. In Italy, the values of $a(g)$ are determined by the National Institute of Geophysics and Vulcanology (INGV), through the analysis of a national catalogue collecting the location and magnitude of the seismic events registered in Italy since AD 1000. 
The indicator \(a(g)\) considered in this study is extracted by the latest maps distributed by INGV, which provide the basis of Italian seismic regulations on constructions. These maps report the median value of the predicted \(a(g)\), as estimated by INGV over a fine grid of points of step 0.02 degrees covering the entire Italian territory \citep{stucchi2004}.
To the best of our knowledge, these maps do not consider possible local amplification effects, which are still under construction in the context of an ongoing national project for the seismic micro-zonation.

Based on these estimates, the MRCI supplies two aggregated summaries of \(a(g)\) at the municipality scale:  $AGMAX\_50$ and $AGMIN\_50$, quantifying the maximum and minimum value of $a(g)$, respectively, taken on the grid points lying within the boundaries of the municipality territorial area. As suggested in \mycitet{Report_2017}, in this paper we shall conservatively consider $AGMAX\_50$ -- in the following, named \(ag[max]\) as in \mycitet{Report_2017} -- to represent the seismic landscape of the country.

\begin{figure}[h]
    \centering\includegraphics[scale = 0.5]{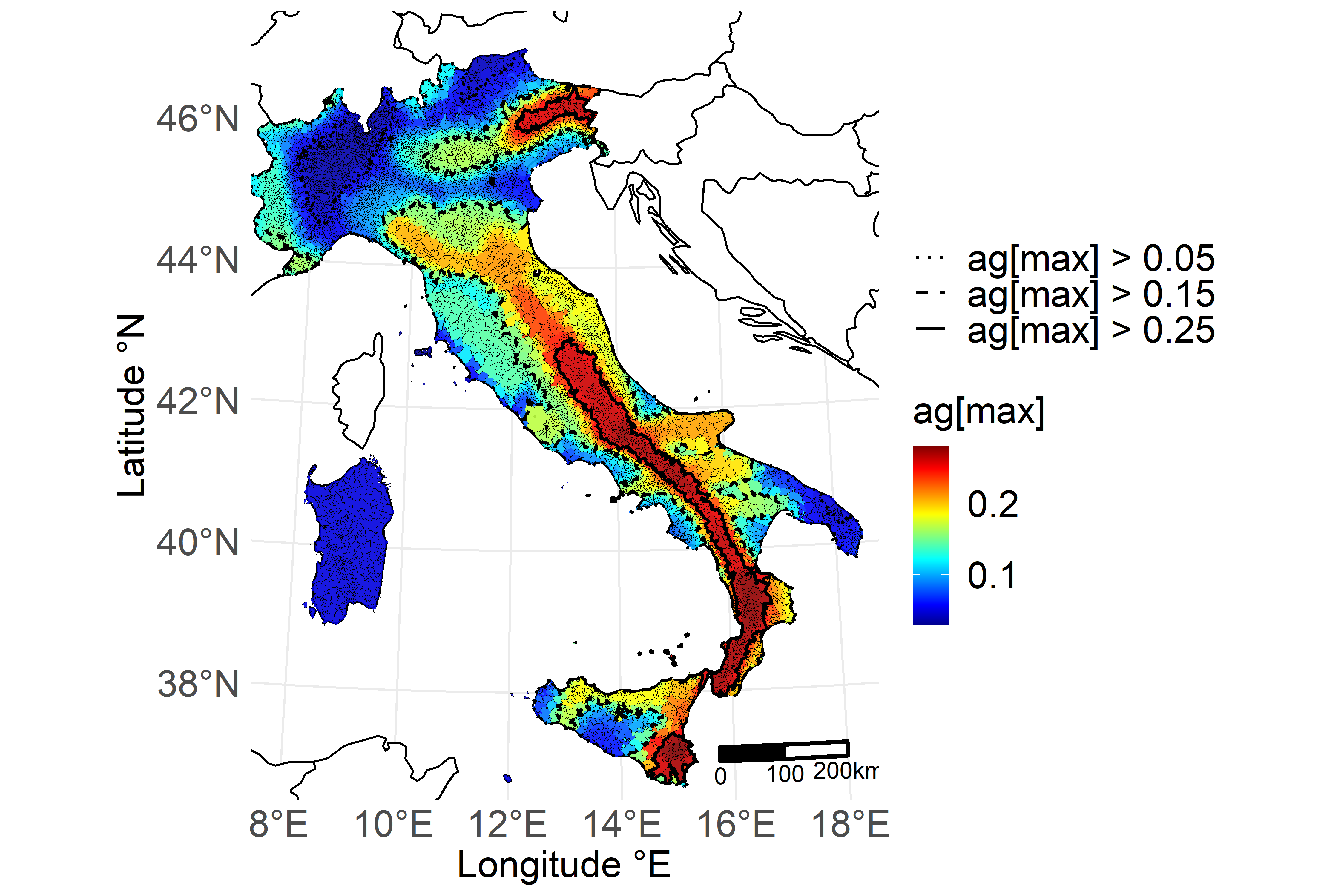}
   \caption{Map of Italy where each municipality has been colored according to \(ag[max].\)}
    \label{fig:a(g)}
\end{figure}

Figure \ref{fig:a(g)} shows the map of Italy where each municipality has been colored according to \(ag[max].\) The figure also shows the map contour lines partitioning the seismic landscape of the country according to four classes of \(ag[max]\), as defined in \mycitet{Report_2017} based on the Italian regulation\footnote{Opcm n. 3519 del 28 aprile 2006: criteri generali per l'individuazione delle zone sismiche e per la formazione e l'aggiornamento degli elenchi delle stesse zone}:
\begin{enumerate}
    \item Low hazard: 0 - 0.05
    \item Moderate hazard: 0.05 - 0.15
    \item Medium hazard: 0.15 - 0.25
    \item High hazard: 0.25 +
\end{enumerate}

\section{The baseline index of social and material vulnerability} \label{sect:IVSM}

For each municipality, the MRCI supplies the value of the Social and Material Vulnerability Index
(\(IVSM\), for {\em Indice di Vulnerabilità Sociale e Materiale}, in Italian), as defined for the Italian census 2011. \(IVSM\) is a scalar meta-index, computed by ISTAT
on the basis of seven different socio-economic indicators: the incidence of population with age between 25 and 64 that is illiterate or without qualification; the incidence of families with at least 6 members; the incidence of single parent families (with age of parent up to 64) over the total of families; the incidence of families with possible welfare poverty; the incidence of population living in severely crowded conditions \footnote{
 This indicator is computed by ISTAT as the ratio (in percentage) between the population being resident in dwelling smaller than 40 sqm with more than 4 occupants, or in 40-59 sqm with more than 5 occupants, or in 60-79 sqm and more than 6 occupants, and the total of the resident population. 
};
the incidence of young people (15-29 years) without occupation; the incidence of families with children with potential economic poverty. The index is an Adjusted Mazziotta-Pareto Index \citep{MazziottaPareto14} which allows for comparison of municipalities across space and time; for more details we refer to \citet{8mila_3,8mila_2}.
The \(IVSM\) index is an estimate of the overall socio-economic vulnerability of a municipality. By construction, all values of \(IVSM\) range in the interval (70,130). High values of the index indicate high vulnerability, while low values indicate low vulnerability; the reference value of 100 corresponds to the value of \(IVSM\) for the entire country in 1991.

\begin{figure}[h]
 \centering\includegraphics[scale = 0.5]{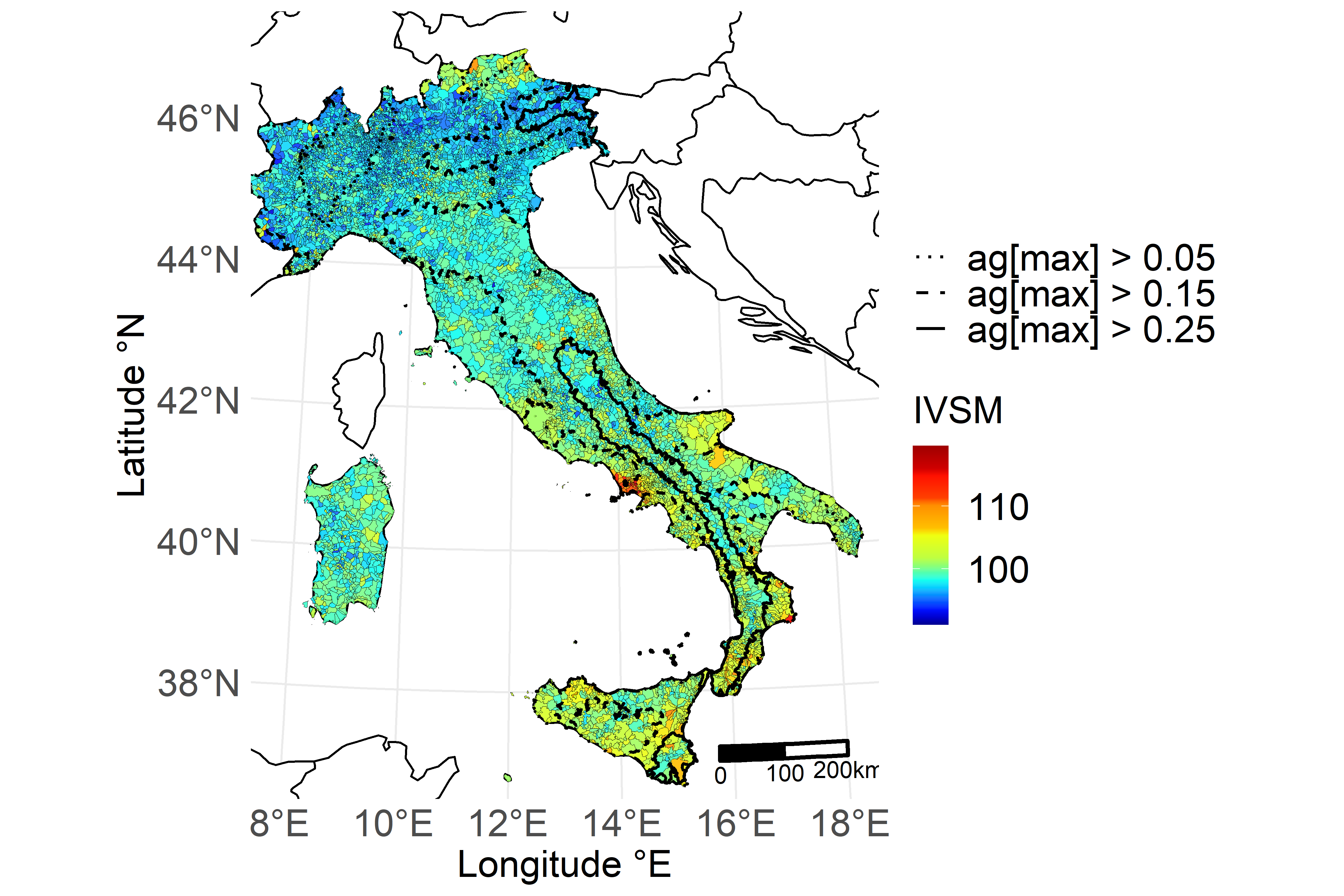}
\caption{Map of Italy where each municipality has been colored according to its \(IVSM\)}
 \label{fig:IVSM_risk}
\end{figure}

We take \(IVSM\) to provide a baseline indication of the inability of the social community (identified as the municipality) to withstand the adverse impacts caused by catastrophic seismic events. %, and to be resilient to its risk. 
Figure \ref{fig:IVSM_risk} shows the value of \(IVSM\) for all the Italian municipalities.  One can observe the spatial heterogeneity of the index across the country, although a general trend is also apparent. In the North, values of \(IVSM\) are typically low, while its highest values are located in the South of the peninsula and in  Sicily. Our first analysis aims at the identification of spatial hot spots for \(IVSM\), i.e., areal clusters where the local spatial correlation of the index is significant. For this purpose we use the Local Indicator of Spatial Association (LISA) first introduced in \citet{Anselin95}. To make the paper self-consistent, a short recap on the measures of spatial association we are going to use is in order.

Consider a spatial domain $D$ partitioned in $n$ areal spatial units. The goal is to measure the spatial association between the values of a variable of interest observed at nearby spatial units. Let $x_i$ be the value of the variable observed at the spatial unit $i$. For instance, $x_i$ is the value of \(IVSM\) for the $i$-th municipality. For any pair $(i,j)$ of spatial units, let \(w_{ij}\) be a weight representing the degree of proximity between \(i\) and \(j.\)  The Moran's  global index of spatial association \citep{Moran}  is defined as
$$
I = \frac{n}{\sum_{i,j} w_{i,j}}\frac{\sum_i \sum_j (x_i - \bar x)w_{ij}(x_j - \bar x )}{\sum_i(x_i - \bar{x})^2},
$$
where \(\bar{x}= (1/n)\sum_i x_i\) is the sample mean of the variable \(x\) taken over all the \(n\) areal spatial units. The analysis of spatial association is commonly based on standardized variables and on row-wise normalized spatial weights, i.e. $\sum_{j=1}^n w_{ij} = 1$ for all \(i.\) In this paper, we will adopt the common practice to set $w_{ij} = 1/n_i$ if there is a common border between the units $i$ and $j$, $n_i$ being the number of neighbouring units of the unit $i$, and \(w_{ij}=0\) otherwise.
Let $\tilde{x}_i$ be the weighted average (named \emph{spatial lag}) of the values of the standardized variable $x$ in the neighbour of the spatial unit \(i.\) Then the expression of  Moran's I statistic is simplified to
\begin{equation}\label{eq:Moran}
I =\frac{1}{n} \sum_i x_i \tilde{x_i}
\end{equation}
which is the correlation coefficient between the values of the variable $x$ and their spatial lags.

The Local Indicator of Spatial Association map (LISA map,  \citet{Anselin95}) identifies spatial units significantly contributing to the global Moran's I index.   Indeed consider the term
\begin{equation}\label{eq:Moran_i}
I_i = x_i \tilde{x_i}
\end{equation}
appearing in the sum defining Moran's I index \eqref{eq:Moran}, as the measure of the local spatial association for the variable \(x.\) On a LISA map the spatial unit \(i\) is highlighted if the local association quantified by the statistic $I_i$ is significantly positive or negative, i.e., the null hypothesis of no autocorrelation is rejected at a certain level of significance. The hypothesis is tested via a permutation scheme \citep{Anselin95,Anselin06} where observed values of \(x\) are randomly reallocated to the areal units, except for the value $x_i$ which is kept fixed. For each random pattern, the statistic \(I_i\) is recomputed; the resulting empirical distribution is used as the reference distribution for the statistic under the null hypothesis of no autocorrelation, i.e. to quantify how extreme is the observed value of \(I_i\) under the complete spatial randomness (CRS, see, e.g., \citet{Cressie93}) assumption. Note that significance of the local autocorrelation depends only on the significance of the spatial lag \(\tilde{x}_i\) (with respect to the null distribution of CSR), since the value $x_i$ is kept fixed.
Significance of the local autocorrelation leads to four possible occurrences: (i) High-High, (ii) Low-Low, (iii) Low-High, (iv) High-Low. High-High (Low-Low) singles out a {\em hot spot}, a spatial unit where the value of $x_i$ is positive (negative) and the spatial lag \(\tilde{x}_i\) is significantly high (low) with respect to the null distribution. On the contrary, Low-High (High-Low) indicates an {\em outlier} spatial unit, where the value of $x_i$ is negative (positive) but the spatial lag \(\tilde{x}_i\) is significantly high (low) with respect to the null distribution.

Hereafter, Moran's I global indexes are computed using the R-package \verb"ape" \citep{ape}. To compute LISA maps along the approach provided in the Python library \verb"PySAL" \citep{Pysal07}, we built an \emph{ad-hoc} R-code, available in the github account indicated in the Introduction. For the LISA maps presented in this work, the level of significance is always set to be \(\alpha = 0.05\).

\begin{figure}
  \begin{subfigure}{7.25cm}
    \centering\includegraphics[width = 7.5cm]{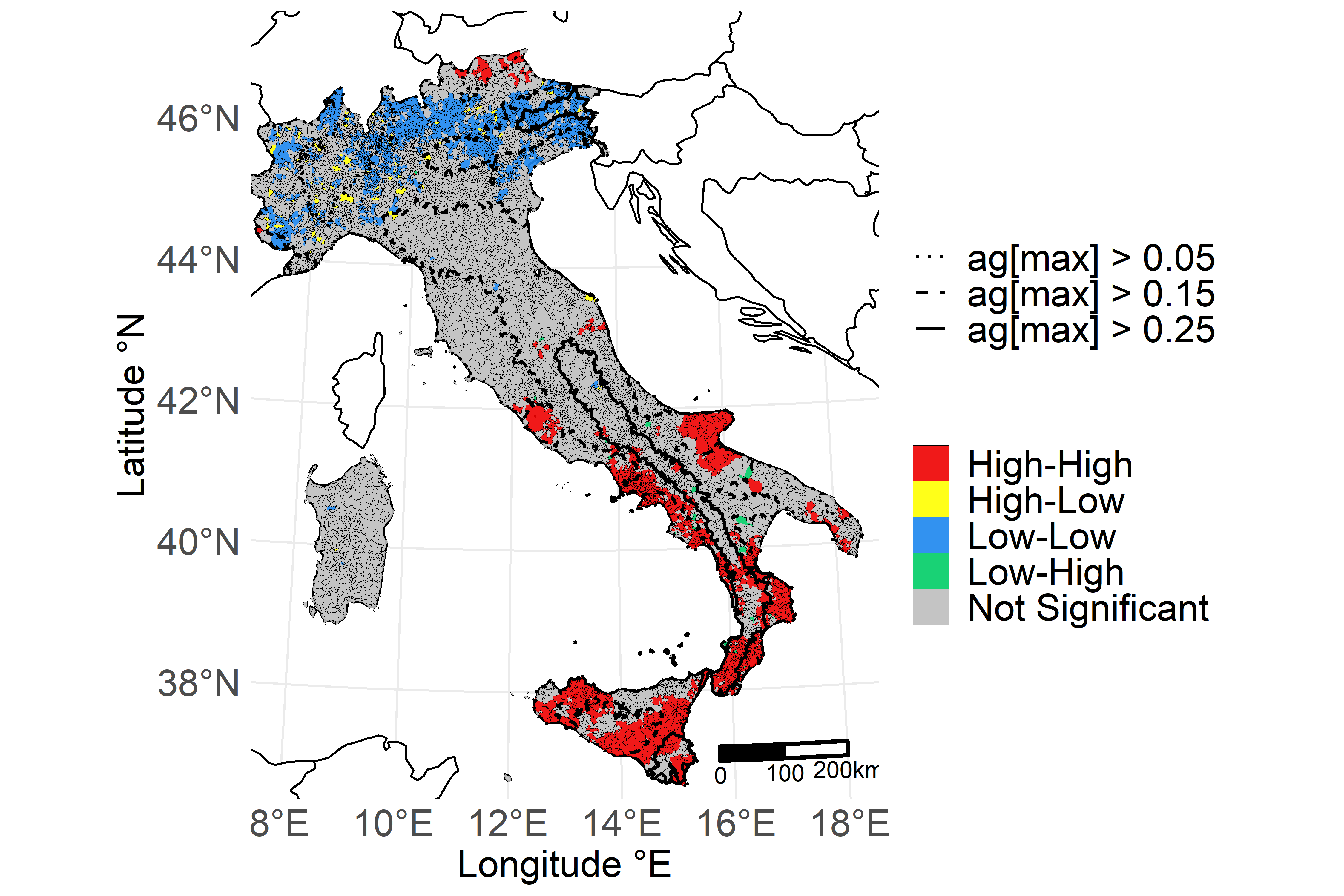}
\caption{ }
  \end{subfigure}
  \begin{subfigure}{7.25cm}
    \centering\includegraphics[width = 7.5cm]{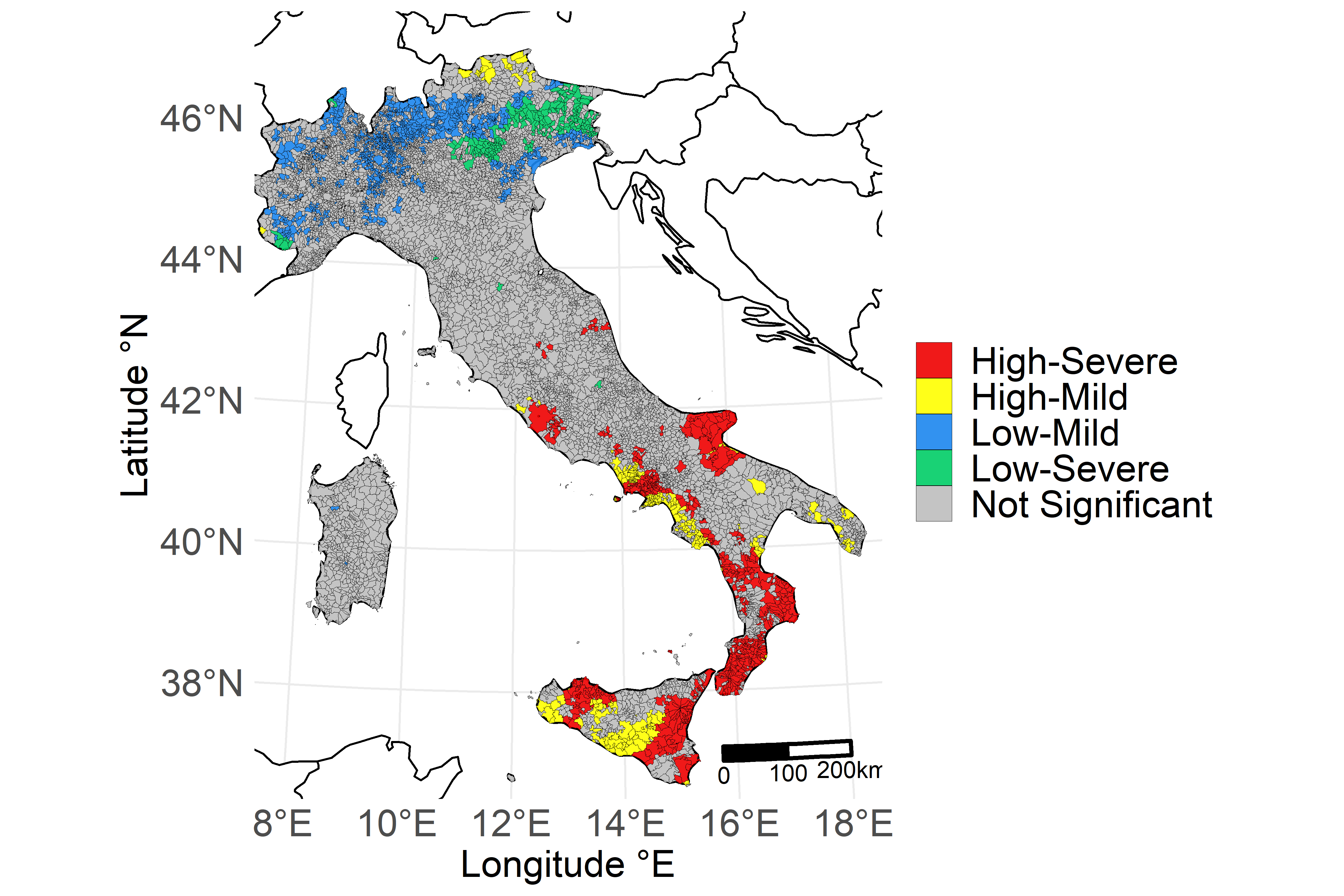}
    \caption{ }
  \end{subfigure}
  \caption{LISA maps for \(IVSM\): (a) is the univariate LISA map for \(IVSM\), (b) the High-High, Low-Low hot-spots of the univariate LISA map stratified by \(ag[max]\) class }
  \label{fig:IVSM_LISA}
\end{figure}

The Moran`s I global statistic measuring the spatial autocorrelation of \(IVSM\) is equal to \(0.644\). Figure \ref{fig:IVSM_LISA}a shows the LISA map of \(IVSM\) for Italy. The High-High red hot spots where \(IVSM\) assumes significant high values are spreading inside the country from the coast in Lazio and in the Southern part of the peninsula. Sizable red areas cover also most of Sicily; notably, the only High-High hot spots in the North of Italy are in Alto Adige. The rest of Northern Italy is characterized by a diffuse presence of blue Low-Low clusters of significantly low values of \(IVSM\).
To enhance interpretation in terms of seismic risk, contour-lines of $ag[max]$ have been superimposed to the LISA map in Figure \ref{fig:IVSM_LISA}a, identifying through dashed lines the regions of high, medium, moderate and low seismic hazard (see Section \ref{sect:seismic_landscape}). Figure \ref{fig:IVSM_LISA}b shows the stratification of the High-High and Low-Low hot spots displayed in Figure \ref{fig:IVSM_LISA}a, according to two macro-classes of $ag[max]$, namely \emph{severe} hazard (high/medium) and \emph{mild} hazard (moderate/low). Here, the spatial clusters colored in red identify municipalities where significantly high values of \(IVSM\) are associated to a severe seismic hazard, their compounded action thus increasing the overall seismic risk.  These are most of the municipalities of the South of Italy already detected in the LISA map of \(IVSM\) because of their high vulnerability. Exceptions are the yellow-colored municipalities of Figure \ref{fig:IVSM_LISA}b, where seismic hazard is mild. Note also the green hot spots in the North, notably in Friuli and Veneto, where a severe seismic hazard is associated with significantly low values of \(IVSM\). For these municipalities, the low value of \(IVSM\) partially compensates the seismic hazard and contributes to the overall decrease of seismic risk.

\section{Population growth and ageing, and social vulnerability }\label{sect:demographics}

Almost a quarter of the Italian population lives in an Inner Area \citep{InnerA_1}. These are rural areas characterized by their distance from the main centers of services (education, health and mobility) and including almost half of the Italian municipalities.
In the last decades, Italian Inner Areas have experienced a pronounced demographic decline and population ageing. These phenomena provide further indications of a fragile community which we believe should be taken into account when evaluating seismic risk and precision policies for reducing it. \(IVSM\) does not include such factors, which therefore deserve a separate analysis.

Daily (and seasonal) population dynamics are complex phenomena, only partially captured by census data, particularly during the daytime. This is also recognized by public authorities, which however agree in finding the global estimate derived from such data to be “acceptable for violent earthquakes that affect large areas”\footnote{
Dipartimento della Protezione Civile (2020) http://www.protezionecivile.gov.it/en/risk-activities/seismic-risk/description 
}. On the other hand, census data offer a reliable source of information having a national coverage with a low selection bias, features which nowadays are hardly found in alternative data sources aiming to model the population dynamics (e.g., data on GPS locations of mobile phones). 

\begin{figure}
  \begin{subfigure}{7.25cm}
    \centering\includegraphics[width = 7.5cm]{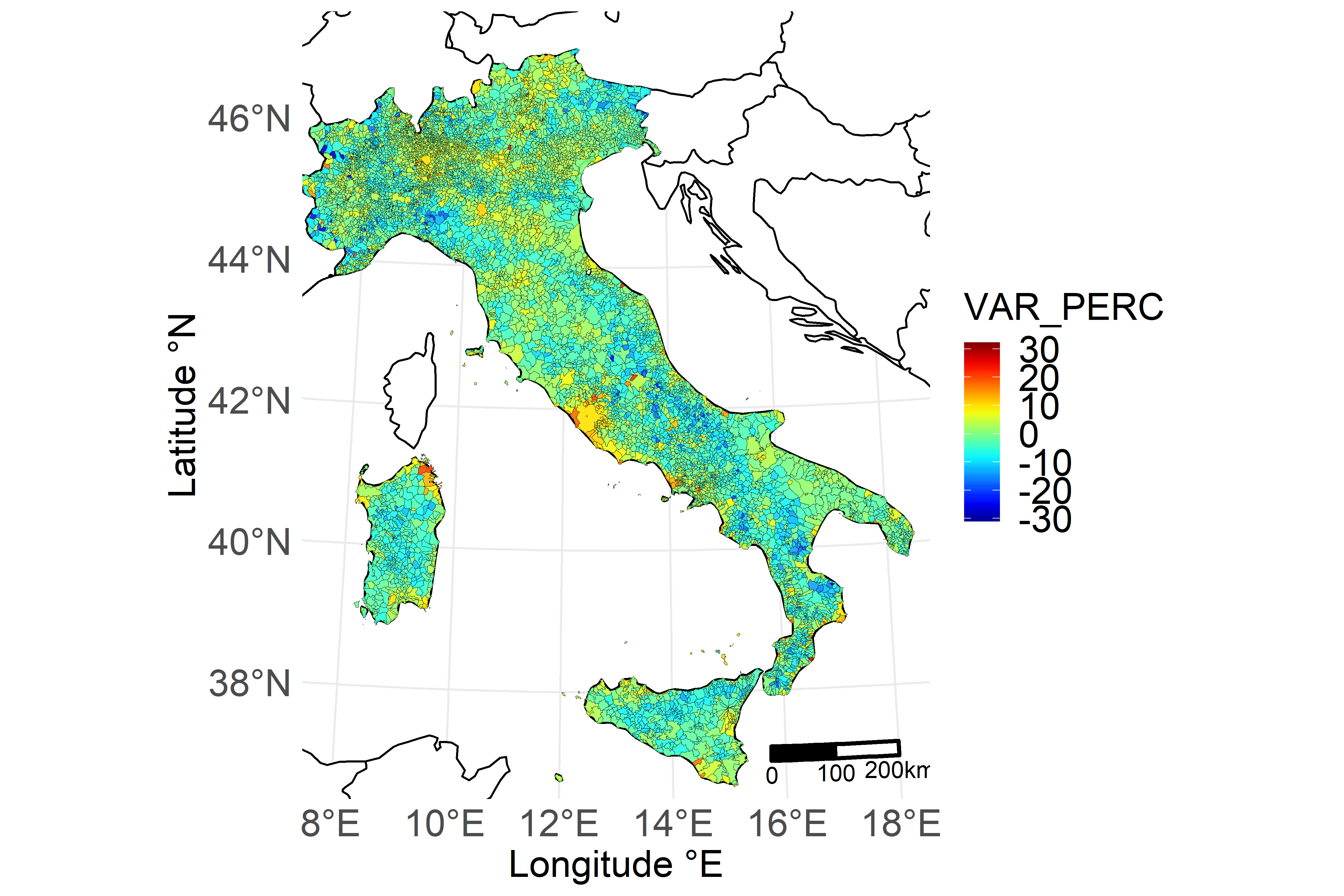}
    \caption{\(VAR\_PERC\) between 2011 and 2018}
  \end{subfigure}
  \begin{subfigure}{7.25cm}
    \centering\includegraphics[width = 7.5cm]{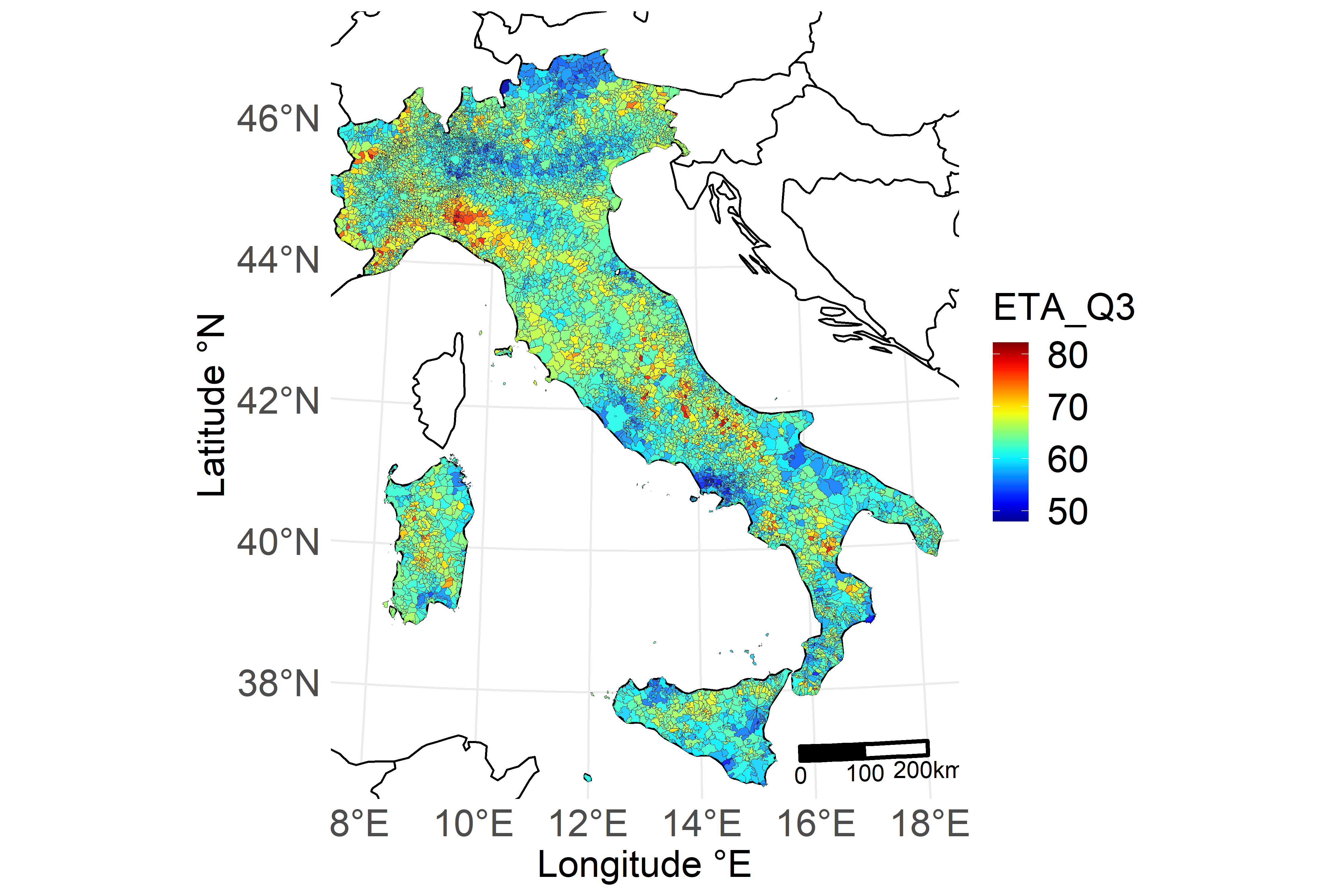}
\caption{\(ETA\_Q3\) in 2018}
  \end{subfigure}
  \caption{Maps of Italy where each municipality has been colored according to its \(VAR\_PERC\) and \(ETA\_Q3\), respectively}
  \label{fig:dep_gr_q3_moi}
\end{figure}

\begin{figure}
  \begin{subfigure}{7.25cm}
    \centering\includegraphics[width = 7.5cm]{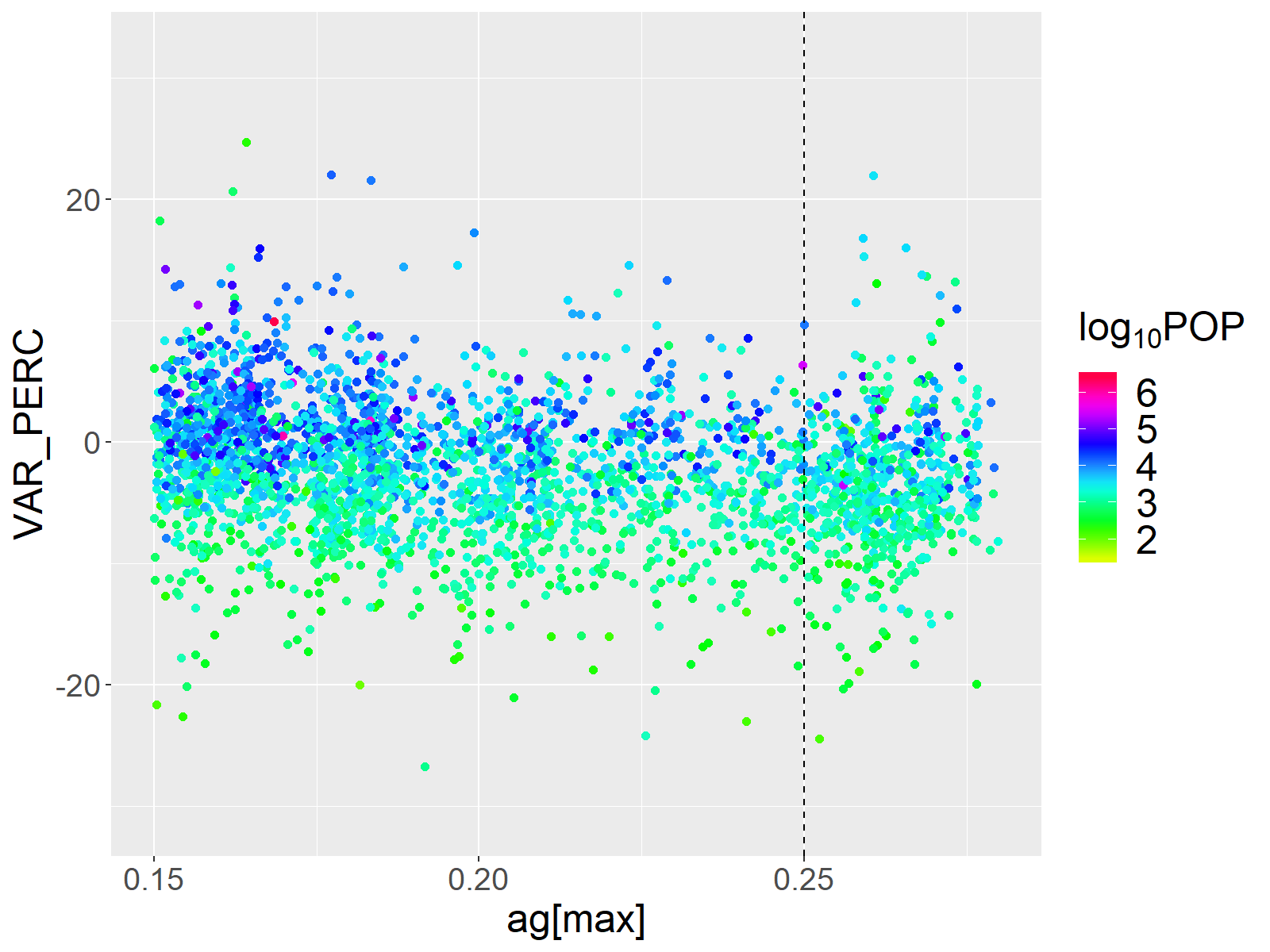}
    \caption{ }
  \end{subfigure}
  \begin{subfigure}{7.25cm}
    \centering\includegraphics[width = 7.5cm]{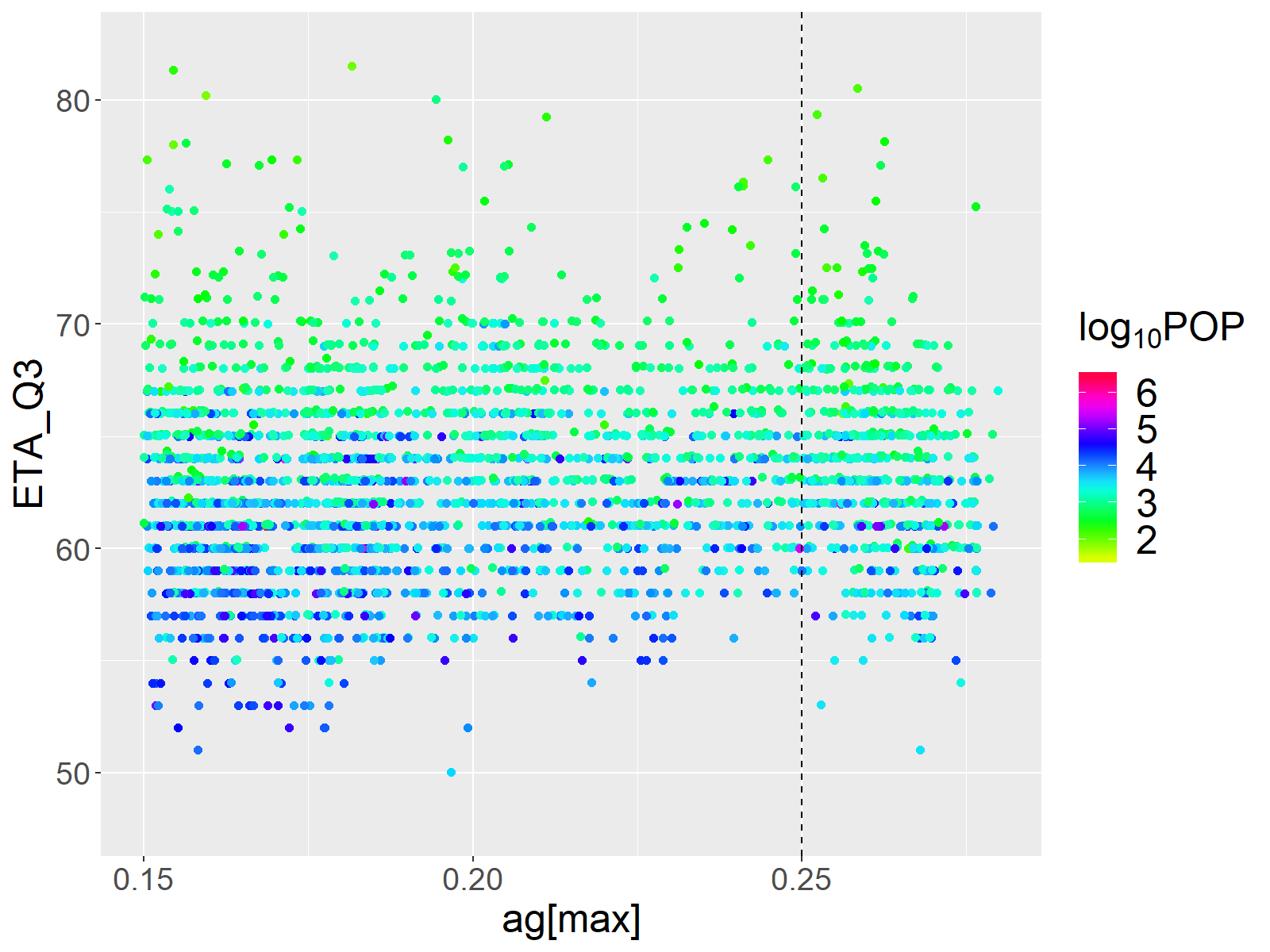}
\caption{ }
  \end{subfigure}
  \caption{Demographic decline and ageing of Italian municipalities projected on the seismic landscape: in map (a) each municipality is represented by its \(ag[max]\) and the value of \(VAR\_PERC\), in map (b) each municipality is represented by its \(ag[max]\) and the value of \(ETA\_Q3\). The color of each municipality corresponds to the \(log_{10}\) of its resident population.}
  \label{fig:dep_gr_q3_sc}
\end{figure}

The MRCI captures the demographic dynamics of each municipality by reporting the percentage variation of the number of its residents in 2018 and 2011, that is the proportion between the difference of number of residents in the two years and the number of residents in 2011, multiplied by 100. In the MRCI this index is called \(VAR\_PERC\).
The age distribution of the residents of a municipality is instead described through its quartiles, as of 2018.
Figure \ref{fig:dep_gr_q3_moi} shows two maps of Italy: on the left, each municipality is colored according to  \(VAR\_PERC\)  whereas on the right the color of each municipality corresponds to the value of the third quartile of the age distribution of its residents, called \(ETA\_Q3\). The values of the global Moran's I statistic for \(VAR\_PERC\) and  \(ETA\_Q3\) are equal to 0.401 and 0.623, respectively. By inspecting the two maps, it is immediately obvious that the two indexes are strongly correlated, the demographic \emph{decline} of the population being correlated with its ageing.  A different perspective on this first qualitative observation, is given by Figure \ref{fig:dep_gr_q3_sc} where, for each municipality, \(VAR\_PERC\) and \(ETA\_Q3\) are plotted against the value of seismic hazard \(ag[max]\), and each municipality is colored according to the \(log_{10}\) of its current resident population size (\(log_{10}(POP)\)).  One may notice that municipalities with a larger population have a higher growth rate and a younger population (i.e., a lower value of \(ETA\_Q3\)).  In general, municipalities with a small number of residents appear older and affected by a pronounced demographic decline. The charts also show that these fragile communities characterize the landscape subject to high seismic hazard.

\begin{figure}
  \begin{subfigure}{7.25cm}
    \centering\includegraphics[width = 7.5cm]{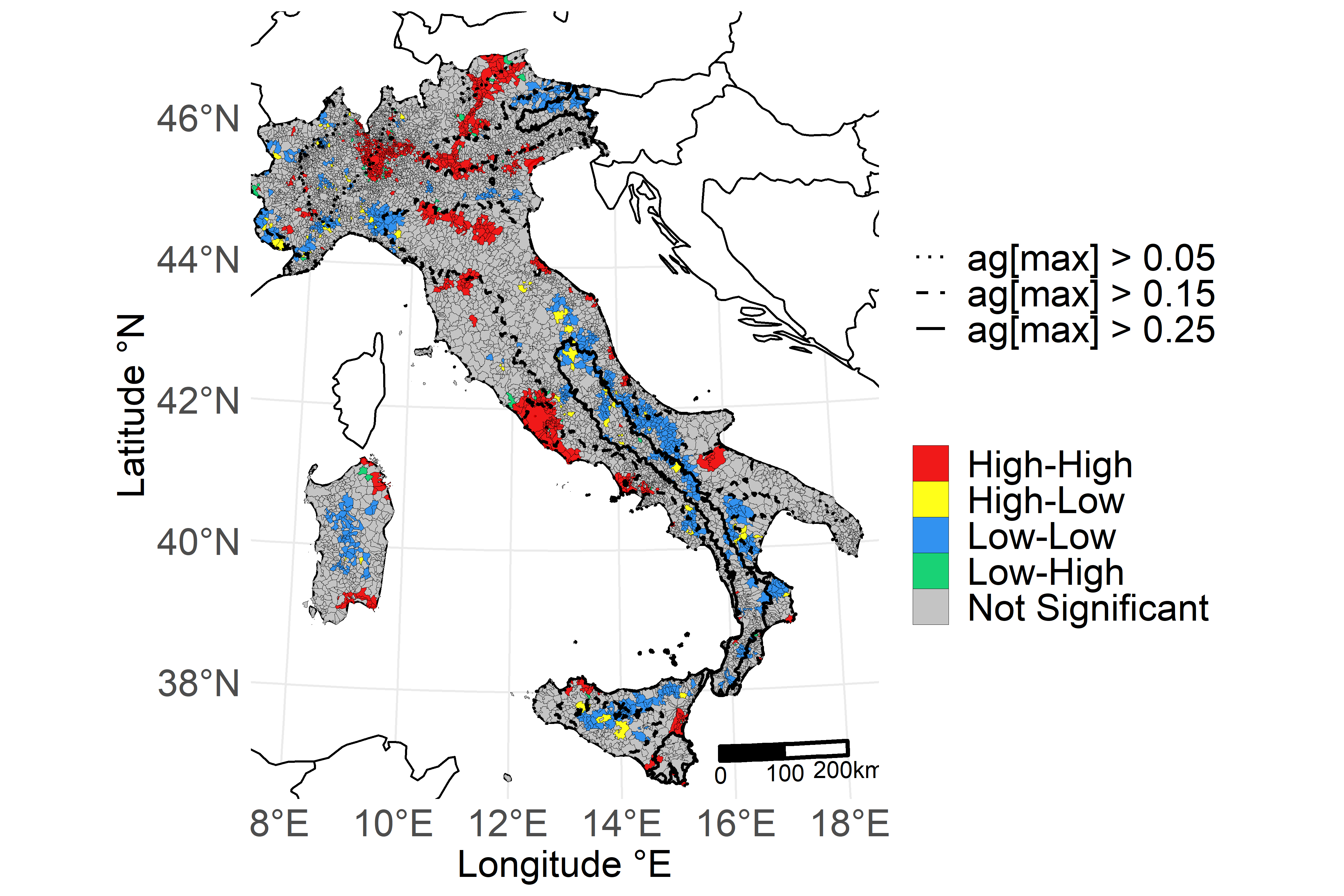}
\caption{ }
  \end{subfigure}
  \begin{subfigure}{7.25cm}
    \centering\includegraphics[width = 7.5cm]{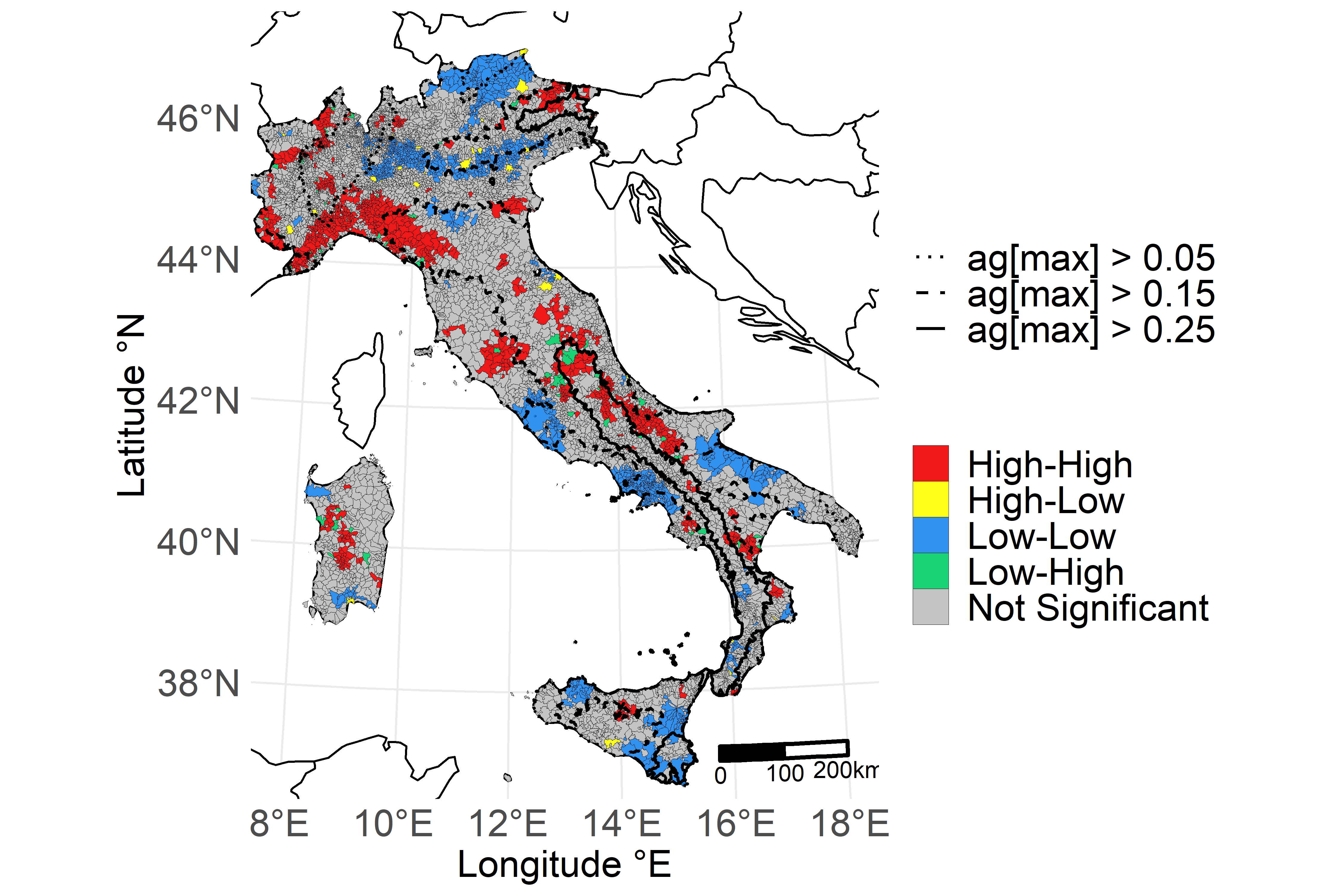}
    \caption{ }
  \end{subfigure}
  \caption{Univariate LISA maps: in map (a) that for \(VAR\_PERC,\) in map (b) that for \(ETA\_Q3\).}
  \label{fig:dep_gr_q3_LISA}
\end{figure}

Aiming at the identification of significant spatial hot spots, we plot the LISA maps of \(VAR\_PERC\) and \(ETA\_Q3\) in Figure \ref{fig:dep_gr_q3_LISA}.
In Figure \ref{fig:dep_gr_q3_LISA}a, the High-High hot spots of \(VAR\_PERC\) indicate a significantly positive growth rate for the population size. They are located in proximity of major urban attractors, in the North East of Italy with the notable addition of areas in Trentino and Alto Adige, in some coastal areas of the South of Italy and the main islands.  Low-Low hot spots are instead located in the inner areas of the country.  The picture is almost replicated in Figure \ref{fig:dep_gr_q3_LISA}b, with colors of opposite sign, where the LISA map for \(ETA\_Q3\) is shown.  Hot spots of significantly high values for \(ETA\_Q3\) indicate areas of ageing populations and are in good agreement with those identifying areas of population decline in Figure \ref{fig:dep_gr_q3_LISA}a. The Low-Low hot spots indicating areas with a younger resident population often correspond to areas characterized by a positive growth in Figure \ref{fig:dep_gr_q3_LISA}a.

\begin{figure}
  \begin{subfigure}{7.25cm}
    \centering\includegraphics[width = 7.5cm]{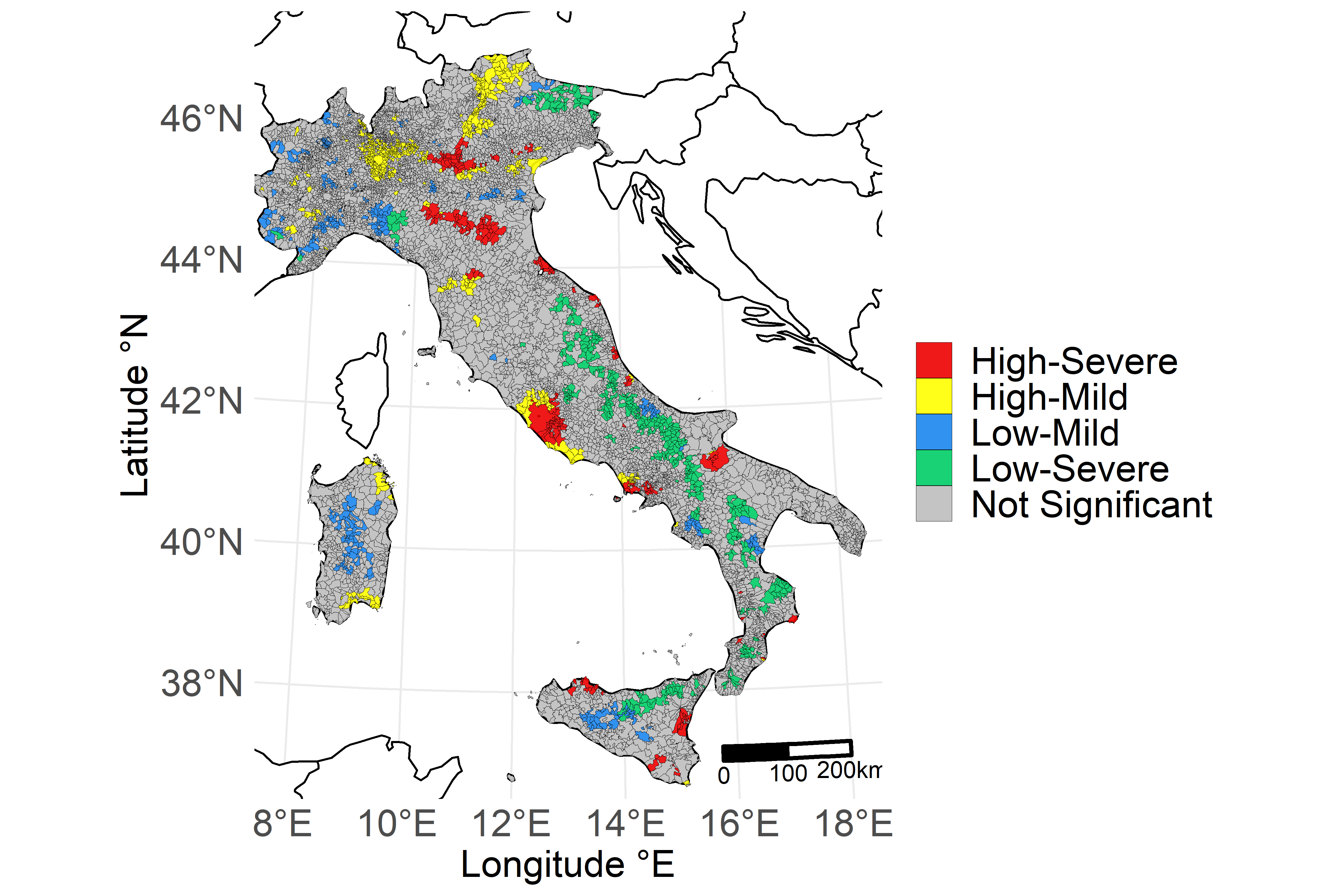}
\caption{}
  \end{subfigure}
  \begin{subfigure}{7.25cm}
    \centering\includegraphics[width = 7.5cm]{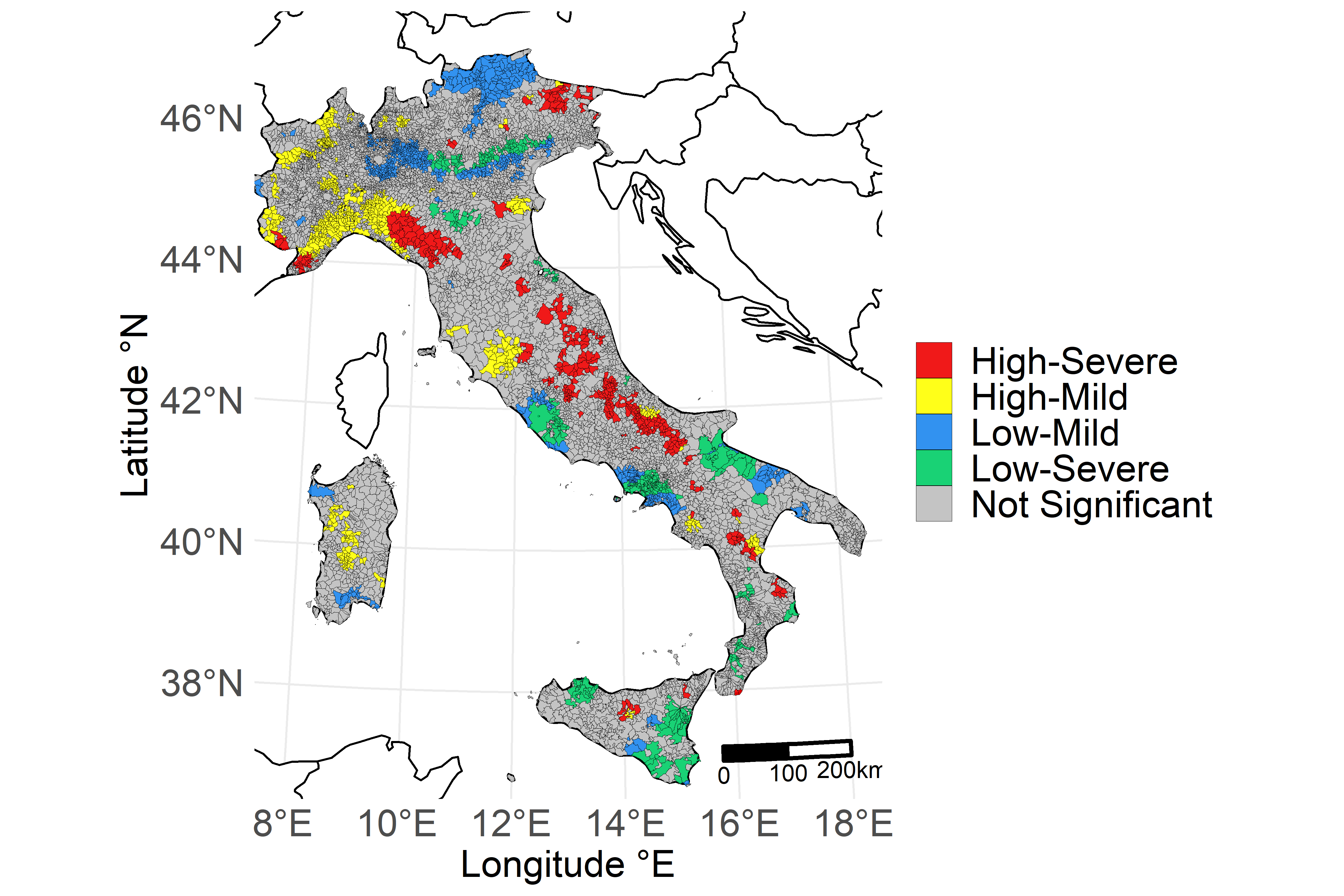}
\caption{}
  \end{subfigure}
  \caption{The High-High, Low-Low hot-spots of univariate LISA maps stratified by \(ag[max]\) class: in map (a) that for \(VAR\_PERC,\) in map (b) that for \(ETA\_Q3\).}
  \label{fig:dep_gr_q3_LISA_strat_ag}
\end{figure}

To project this analysis on the seismic landscape, we plot in Figure  \ref{fig:dep_gr_q3_LISA_strat_ag} the LISA maps of Figure \ref{fig:dep_gr_q3_LISA} stratified according to the two macro-classes of $ag[max]$ identifying severe and mild seismic hazard. Areas where a severe seismic hazard (high or medium values of \(ag[max]\)) are associated with a negative growth rate of the population are of particular concern for seismic risk prevention and reduction; these are the green hot spots in Figure \ref{fig:dep_gr_q3_LISA}a. The same concern applies to the red hot spots in Figure \ref{fig:dep_gr_q3_LISA}b, identifying areas where a severe seismic hazard is significantly associated to an ageing population. For many of these areas, particularly those belonging to the inner areas of the Appenine, there is a complete overlap with the green hot spots in Figure \ref{fig:dep_gr_q3_LISA}a, thus indicating communities where old age and a decreasing population add up to the their fragility in the face of a relevant seismic hazard.

\section{The building stock age}\label{sect:building_stock}
Human communities live in buildings, and building vulnerability is the first and most important component of any assessment of seismic risk. In this vein, the main efforts of the Casa Italia project was to envision the quantitative basis to design efficient and effective precision policies incentivizing the reduction of vulnerability of the heterogenous building stocks of the Italian municipalities \mycitep{Report_2017}, thus promoting the safety of their communities against seismic risk. We argue that building stock vulnerability, as measured by its age distribution, should also be considered when evaluating the social and material vulnerability of a community. An old building stock -- in Italy, often composed of historical buildings which cannot be demolished and reconstructed -- requires large interventions for enhancing its quality against seismic shaking. The inability of a community to put in place such preventive actions, favoured by its demographic decline and ageing,  adds to its overall fragility.
In this regard, we note that, although gentrification characterizes some developments stimulated by national strategies aimed at the economic growth of old towns this phenomenon still appears to be of minor entity with respect to the pervasive dynamics of desertion which are affecting large inland areas of Italy. Indeed, desertion is still a motivating factor for the national program for Inland Areas \footnote{
Department of Public Administration, National strategy for Inland Areas,  
http://www.pongovernance1420.gov.it/en/Project/inland-areas/
}, established by the Italian Government to develop and valorize those areas. Thus, in the context of our study, an indicator of the age of the building stock allows us to indirectly consider the overall dynamics of desertion of Inland Areas, besides providing a proxy of the overall seismic vulnerability of the building stock itself. We however remark that the aim of the present analysis is not to quantify the \emph{physical} vulnerability of the buildings but rather include this indirect signal into a quantification of the vulnerability of the community. In fact, additional sources of information for the \emph{physical} vulnerability of the building stock would be provided by, e.g., the type of material (masonry, wood, concrete), number of floors, state of conservation etc. (see, \citet{d2002integrated, lang2002seismic,dolce2006vulnerability,salvucciIstat, maio2015seismic}).

\begin{figure}
\centering
\includegraphics[scale = 0.5]{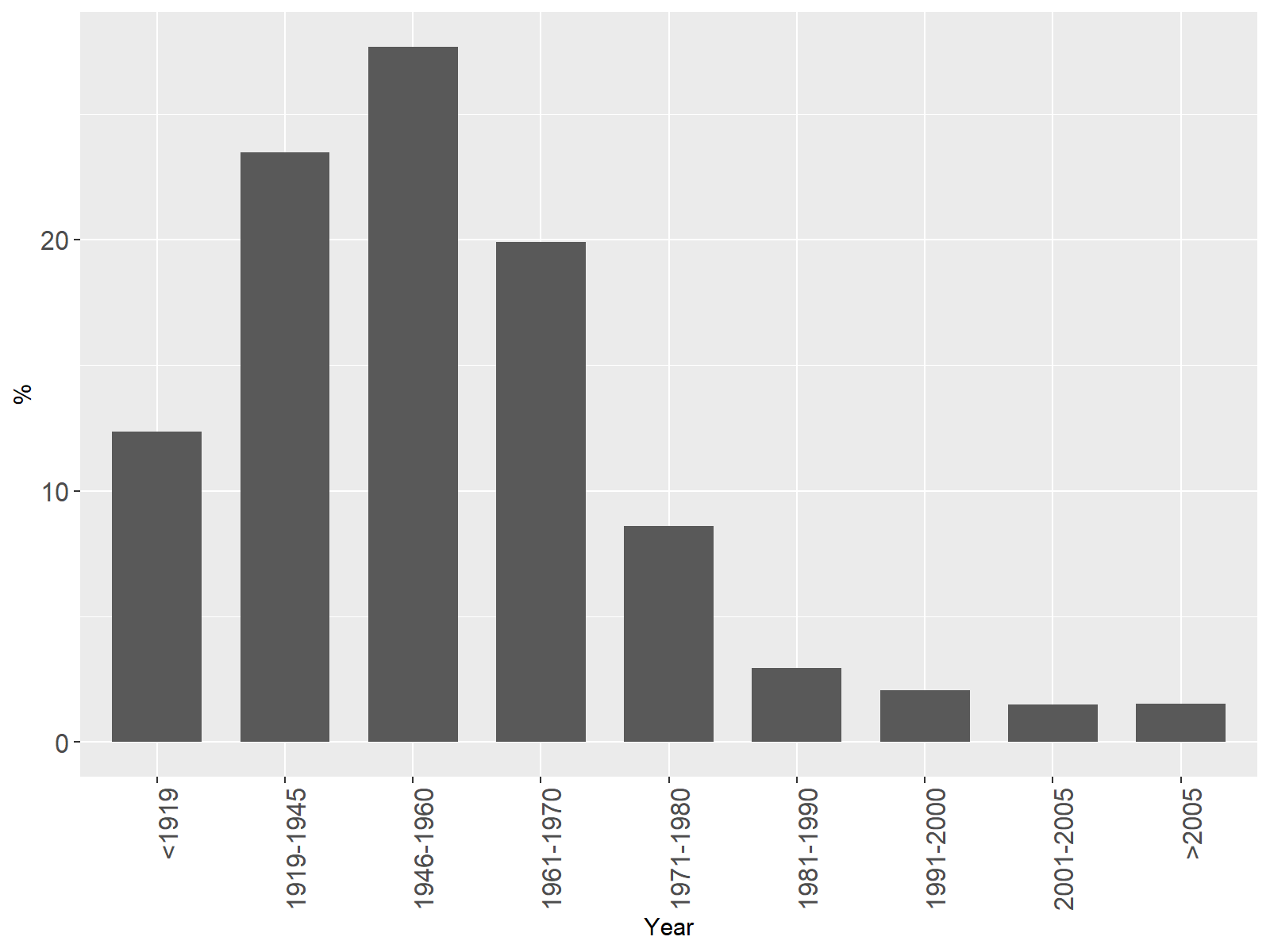}
\caption{Building stock distribution for the municipality of Milan}
\label{fig:build_milano}
\end{figure}

For each municipality, the MRCI summarizes the age of its building stock by reporting the relative frequencies of buildings constructed in each of the following nine disjoint time classes: before 1919; 1919-1945; 1946-1960; 1961-1970; 1971-1980; 1981-1990; 1991-2000; 2001-2005; after 2005. See, for instance, Figure \ref{fig:build_milano} for the distribution of time-of-construction of the building stock of Milan. Time-of-construction is an important indicator of building vulnerability since anti-seismic regulations, enforcing higher standards for building quality against earthquake shaking, have been introduced in Italy between 1981 and 1984 and then again in 2003: see \mycitet{Report_2017}.

For the analysis of the variability of the time-of-construction distributions of the building stock of the Italian municipalities, we note that these are compositional data which belong to the simplex of $R^9$. Indeed, the time-of-construction distribution for the buildings of the $i$-th municipality is a vector of \(p=9\) positive components $\underline{x}_i = (x_1, ..., x_p)'$
summing up to 1. The vector $\underline{x}_i$ belongs to the simplex $S^p$, a subset of $R^p$:

$$
S ^ { p } = \left\{ \underline { x } = \left( x _ { 1 } , \ldots , x _ { p } \right) ^ { \prime } \in R^p : x _ { j } > 0 \,\mbox{for}\; j=1,...,p, \; \; \sum _ { j = 1 } ^ { p } x _ { j } = 1 \right\}.
$$

\citet{Aitchison82} firstly proposed the use of a log-ratio approach to the analysis of compositional data, defining a set of principles that such analyses should fulfill (see \citet{Pawlowsky-Glahn, CoDA_book, CoDA_book_2}). These principles led to the definition of a specific geometry for \(S^p\), named \emph{Aitchison geometry}, which is built upon two main operations, perturbation and powering defined, for \(\underline{x}, \underline{y} \in S^p\), as
\[
(\underline{x} \oplus \underline{y})_j=\frac{x_jy_j}{\sum_{k=1}^px_ky_k}; \;(\alpha\odot\underline{x})_j=\frac{x_j^\alpha}{\sum_{k=1}^px_k^\alpha},\quad \]
and an inner product
\[
\langle \underline{x}, \underline{y} \rangle = \frac{1}{2p} \sum_{j=1}^p\sum_{k=1}^p \ln{\frac{x_j}{x_k}} \ln{\frac{y_j}{y_k}},\]
implying a norm and a distance within the space. Note that the neutral element of perturbation is the uniform composition, i.e., $\underline{0}_{\oplus} = (1/p,...,1/p)^\prime$.

The simplex \(S^p\) endowed with the Aitchison's geometry is a finite-dimensional Hilbert space, which is isometrically isomorphic to the linear space \(M\subset R^p\), of dimension \(p-1\), consisting of those vectors of \(R^p\) whose components sum up to 0.  The isomorphism is, for instance, provided by the \emph{centered log-ratio transformation} ($clr$) defined, for \(\underline{x} \in S^p,\) as
\begin{equation}\label{eq:clr}
clr(\underline{x})=\left[\ln \left(\frac{x_1}{(\prod^p_{j = 1} x_j)^{1/p}}\right),..., \log \left(\frac{x_p}{(\prod^p_{j = 1} x_j)^{1/p}}\right)\right]^\prime.
\end{equation}

Since building stock data are constrained objects, the overall effect of the seismic hazard factor, whose levels are the four classes listed in Section \ref{sect:seismic_landscape}, on the distribution of building stock compositions is here assessed through a \emph{compositional} one-way ANOVA. For the time-of-construction composition $\underline{x}_{is}$ of the building stock of the $i$-th municipality in the $s$-th class of seismic hazard, we thus consider the model
\begin{equation}\label{eq:comp-anova}
\underline{x}_{is} = \underline{\mu} \oplus \underline{\tau}_s \oplus \underline{\epsilon}_{is},
\end{equation}
with $\underline{\mu}$ the overall mean, $\underline{\tau}_s$ the effect of the $s$-th level of the seismic hazard factor, and $\underline{\epsilon}_{is}$ zero-mean i.i.d. errors. To test the significance of the $\underline{\tau}_s$,
we set the hypothesis
\begin{equation}\label{eq:ANOVA-test}
H_0:\, \underline{\tau}_1 = ... = \underline{\tau}_4=\underline{0}_{\oplus}\,\, \textrm{ against }\,\, H_1:\,\mbox{at least one}\; \underline{\tau}_s \neq \underline{0}_{\oplus},
\end{equation}
and we perform a permutational ANOVA test, that allows coping with the non-Gaussianity of the errors \citep{Permanova_10, Permanova_17}. From the computational standpoint, the test is run by first transforming the data thorough the $clr$ function defined in \eqref{eq:clr}, which yields a set of (constrained) data $\{clr(\underline{x}_{is}), i=1,...,n_s,\, s=1,...,4\}$ in the linear space $M \subset R^p$, where \(n_{s}\) stands for the number of municipalities belonging to the \(s\)-th level class of seismic hazard. Due to the isometric nature of the $clr$ transformation, the following model in $R^p$ for the $clr(\underline{x}_{is})$'s is consistent with \eqref{eq:comp-anova}
\begin{equation}\label{eq:clr-anova}
clr(\underline{x}_{is}) = clr(\underline{\mu}) + clr(\underline{\tau}_s) + clr(\underline{\epsilon}_{is}).
\end{equation}
On the transformed model, the hypotheses \eqref{eq:ANOVA-test} read
\begin{equation}\label{eq:ANOVA-test-clr}
H_0:\, clr(\underline{\tau}_1) = ... = clr(\underline{\tau}_4)= \underline{0}\,\, \textrm{ against }\,\, H_1:\, \mbox{at least one}\; clr(\underline{\tau}_s) \neq \underline{0}.
\end{equation}
The test is run using the pseudo-F statistic
\[
F = \frac{(SS_T-SS_W)/(4 - 1)}{SS_W/(n-4)},
\]
where $SS_T$ is the total sum-of-squares and $SS_W$ is the within groups sum-of-squares:
\[
\begin{split}
   SS_T & = \frac{1}{n}\sum_{i = 1}^{n - 1}\sum_{j = i +1}^{n} d^2_{ij}, \\
   SS_W & = \frac{1}{n}\sum_{i = 1}^{n - 1}\sum_{j = i +1}^{n} d^2_{ij}\delta_{ij}.
\end{split}
\]
Here $d_{ij}$ indicates the Euclidean distance between clr-transformed observations $i$, $j,$ while $\delta_{ij} = 1$ if observations \(i\) and \(j\) belong to the same group and $0$ otherwise \citep{Anderson01,Permanova_17}.

The distribution of \(F\) under \(H_0\) is obtained by permuting the transformed data, as advocated in \citet{Permanova_10}. 
Under \(H_0,\) the pseudo-\(F\) statistic  exceeds its observed value $F_0= 54.35$ in the sample with an estimated frequency equal to $0.001$. The routines that implement the test performed with the R package \verb"vegan" \citep{vegan} are freely available in the github account indicated in the Introduction; they have been run by setting the number of permutations equal to  1000.
The result supports the claim that the effect of the seismic hazard factor on the distribution of the time-of-construction compositions is significant.
This shows that the mean distribution of the building stock is significantly different in the different areas identified by the seismic hazard indicators. A closer look at the data suggests that the areas of higher seismic hazard appear associated with age-distributions more concentrated on older buildings, which contribute to make these areas more prone to a higher seismic risk.

Indeed, we further investigate this issue, by exploring the spatial distribution of the time-of-construction compositions of the building stock through effective compositional summaries. We reduce the dimensionality of the of the time-of-construction compositions by eliciting the main direction of their variability via a Compositional Principal Component Analysis \citep{Aitchison83}  performed with the R package \verb"compositions" \citep{compositions}.

\begin{figure}
  \begin{subfigure}{7.25cm}
    \centering\includegraphics[width=7.5cm]{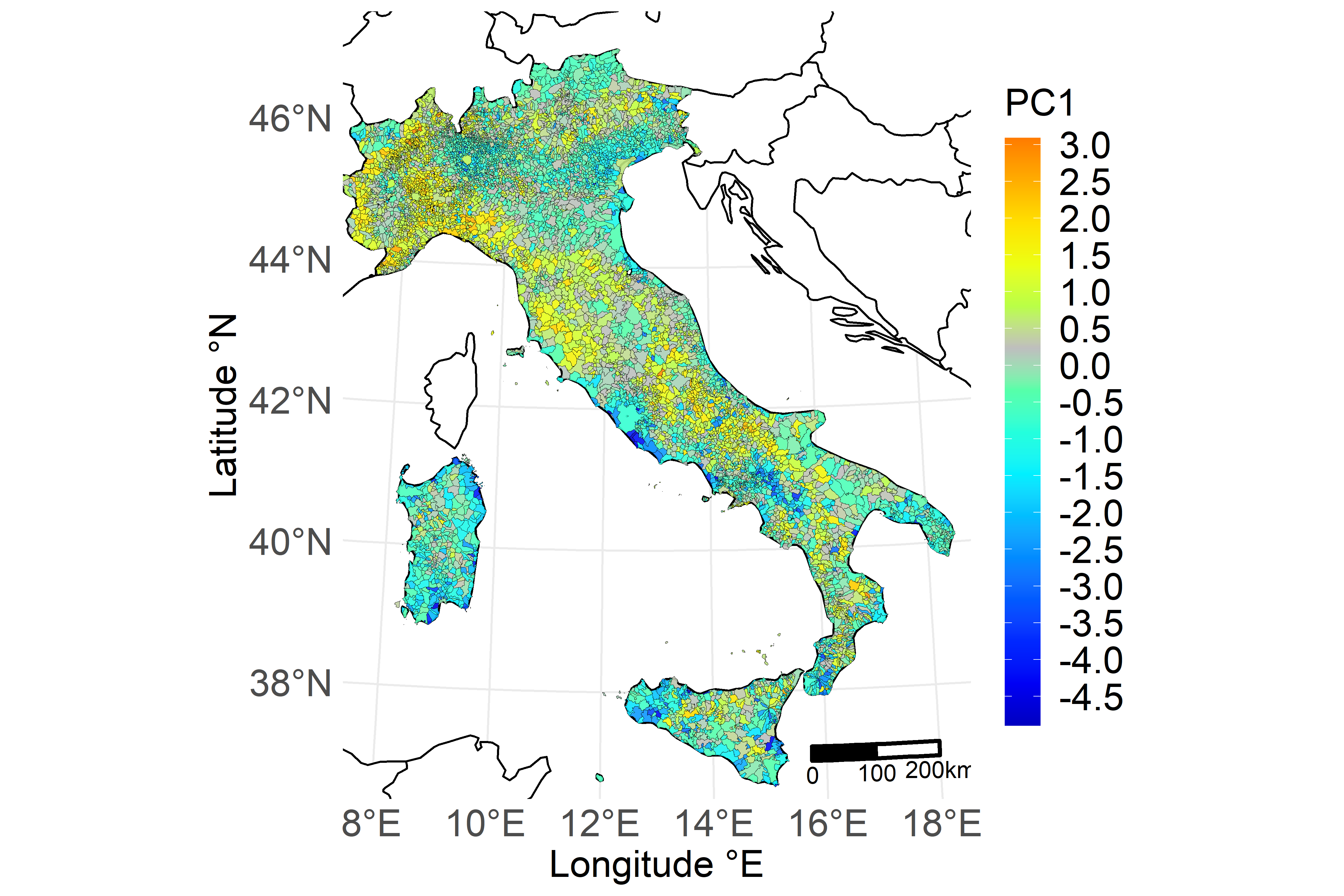}
    \caption{ }
  \end{subfigure}
  \begin{subfigure}{7.25cm}
    \centering\includegraphics[width=6.6 cm]{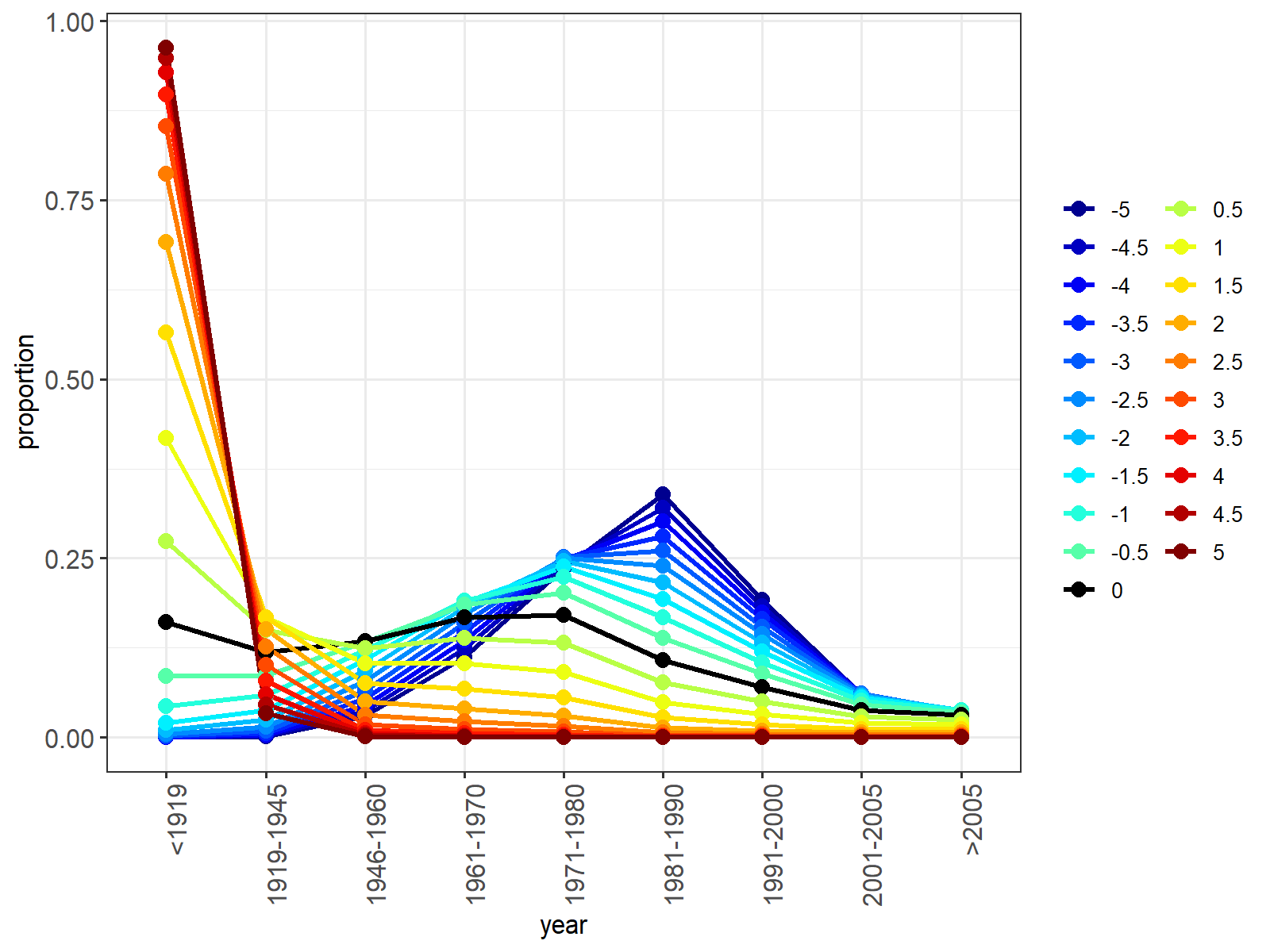}
    \caption{ }
  \end{subfigure}
  \caption{Building stock age: in map (a) each municipality has been colored according to the \(PC1\) score of the time-of-construction distribution of its building stock, map (b) reports the guide chart associating scores with time-of-construction distributions.}
  \label{fig:Building_PC1}
\end{figure}

Figure \ref{fig:Building_PC1} shows the first principal component scores - a variable called \(PC1\) in the following - and projections, this principal component explaining  the $45\%$  of the variability of the time-of-construction compositions.
The guide chart in Figure \ref{fig:Building_PC1}b reports in black the average building stock composition for Italian municipalities, computed according to the Aitchison geometry in \(S^9\). This barycentric distribution has score equal to 0 in the direction of the first principal component. The colored distributions appearing in the same chart correspond to distributions lying on the first principal component direction and identified by a score ranging from \(-5 \sqrt{\lambda_1}\) to \(+5\sqrt{\lambda_1}\), $\sqrt{\lambda_1} = 1.67$ being the standard deviation of the data scores along the first principal component. Low score values (cold colors) indicate time-of-construction distributions concentrated on more recent years (i.e. a young building stock) whereas high score values (hot colors) correspond to time-of-construction distributions more concentrated in the long past (i.e. an old building stock). This chart guides the reading of the Italian map in Figure \ref{fig:Building_PC1}a, where each municipality is colored according to \(PC1.\)  Colors indicate the correspondence between scores, as represented on the Italian map in Figure \ref{fig:Building_PC1}a, and time-of-construction distributions plotted in the guide chart in Figure \ref{fig:Building_PC1}b. Municipalities with a cold color have a younger building stock, those colored with hot colors an older one.

\begin{figure}
\centering
 \begin{subfigure}{7.25cm}
    \includegraphics[width = 7.5cm]{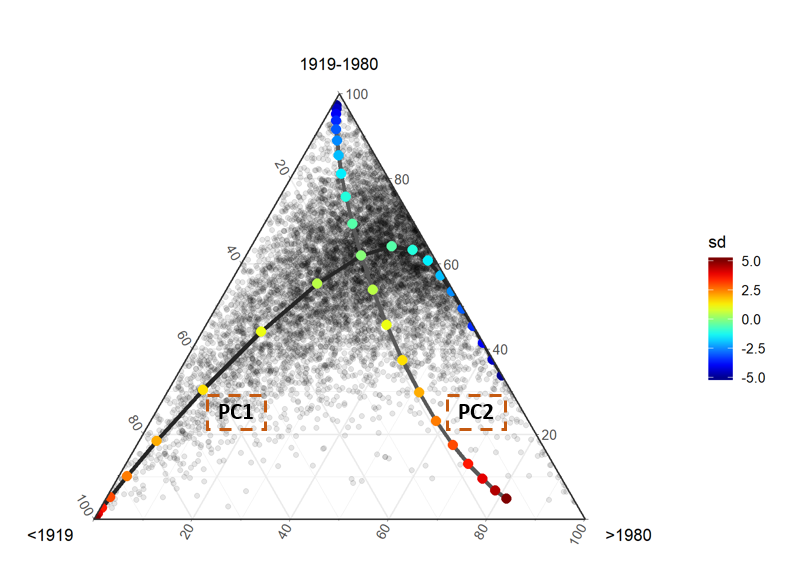}
\caption{ }
  \end{subfigure}
  \begin{subfigure}{7.25cm}
    \centering
    \includegraphics[width = 7.5cm]{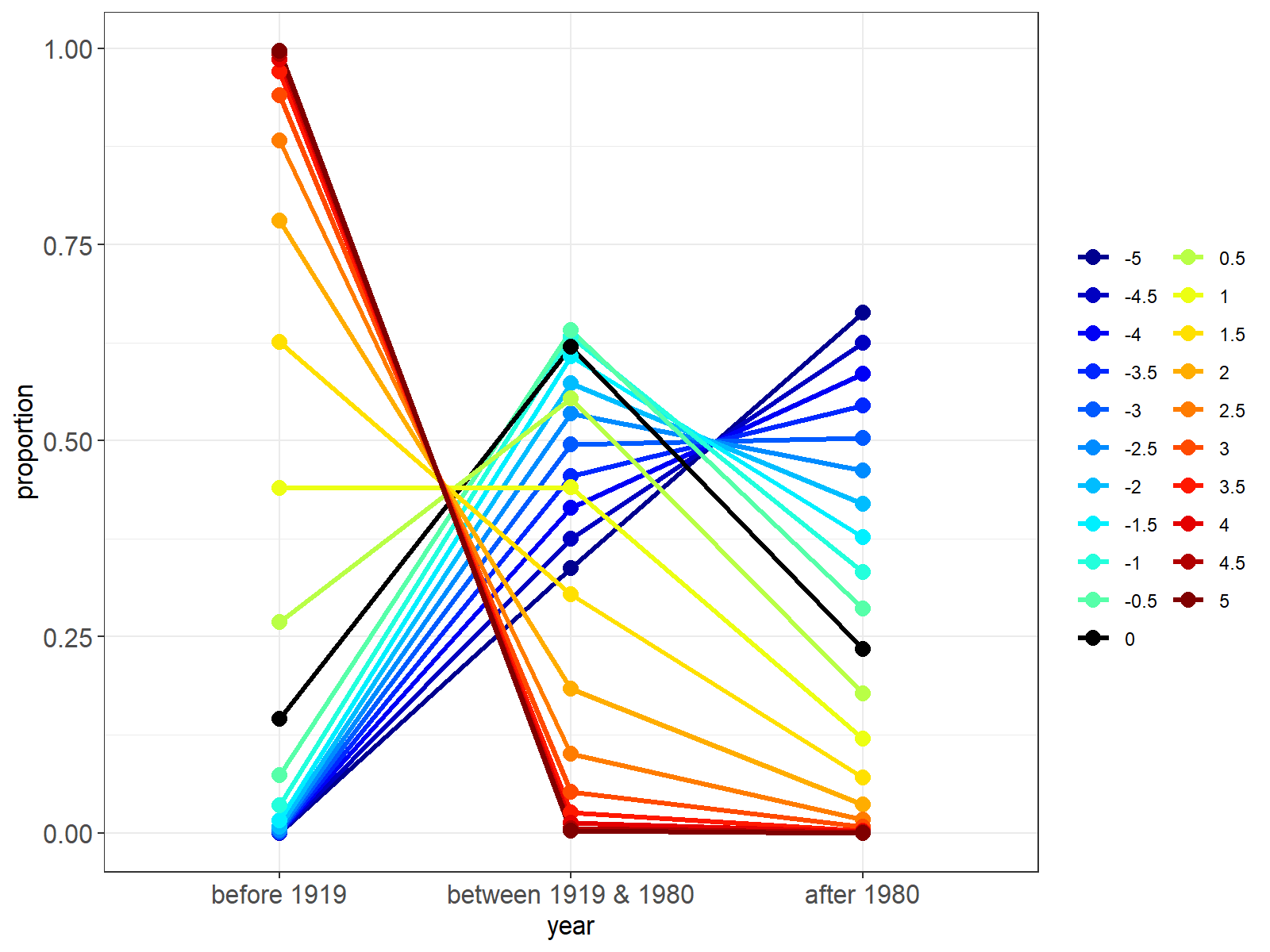}
\caption{ }
  \end{subfigure}
  \begin{subfigure}{7.25cm}
  \centering    \includegraphics[width = 7.5cm]{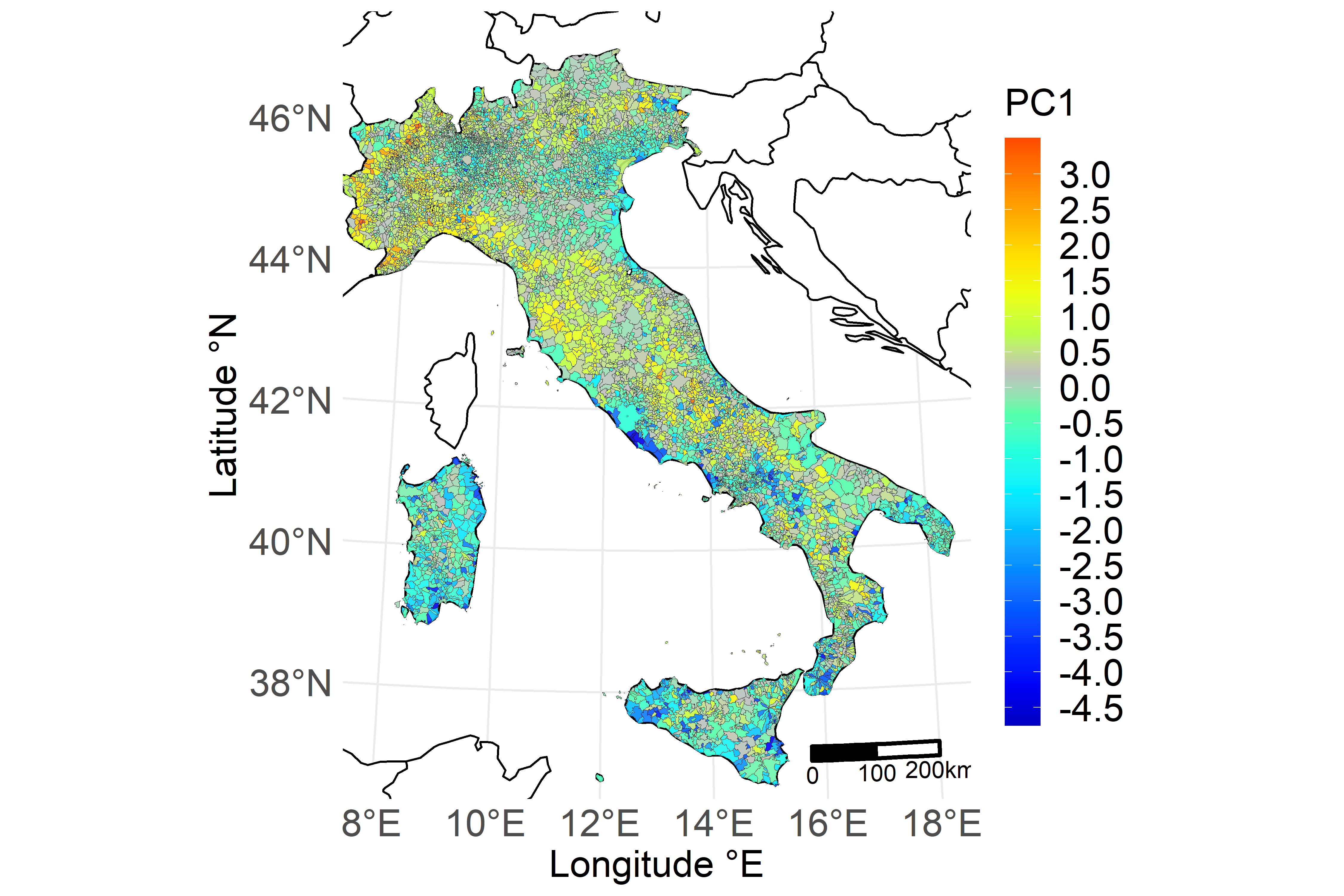}
\caption{ }
  \end{subfigure}

  \caption{Building stock age: each point in the simplex shown in map (a) represents the ternary distribution of the time-of-construction of the building stock of an Italian municipality. The basis formed by the two principal components is shown; \(PC1\) and \(PC2\) cross at the average ternary distribution. The colors of the scores identifying points on the \(PC1\) axis correspond to the colors of the distributions appearing in the guide chart (b) on the right. In map (c) each municipality is colored according to the \(PC1\) score of the ternary distribution of the time-of-construction of its building stock.}
  \label{fig:ternary}
\end{figure}

To emphasize the impact of anti-seismic regulations, we aggregate the time classes defining the time-of-construction distributions in three macro time intervals: before 1919 (very old buildings), between 1919 and 1980 (more recent buildings, constructed before the anti-seismic regulations), after 1980 (buildings constructed while or after anti-seismic regulations were introduced in Italy). These time-of-construction ternary distributions, one for each Italian municipality,  can be represented as points belonging to the simplex \(S^3\). The data cloud appears in Figure \ref{fig:ternary}a. Reported on the simplex are also the directions of their two principal components, which define a basis for \(S^3\) and which intersect at the barycentric distribution, corresponding to the average ternary distribution of time-of-construction for Italian municipalities. The first principal component (\(PC1\)) explains \(86\%\) of the data variability; by moving along the \(PC1\) direction, passing from positive (hot) scores to negative (cold) scores, one visits time-of-construction ternary distributions first concentrated on the ``before 1919'' class and then more and more concentrated on the ``after 1980'' class, as is also shown in the guide chart in Figure \ref{fig:ternary}b.
Figure \ref{fig:ternary}c represents each Italian municipality colored according to the score of its time-of-construction ternary distribution on the \(PC1\) represented in Figure \ref{fig:ternary}a and \ref{fig:ternary}b. One could consider this map of Italy as a coarse summary of the  map in Figure \ref{fig:Building_PC1}. Notice, in particular, the areas of the country where the building stocks are of more recent time of construction (colored in blue); among the obvious metropolitan areas, the Irpinia region and some areas of Friuli stand out, both regions having been vastly reconstructed after the tragic earthquakes of 1976 and of 1980, which prompted the Italian Parliament to pass anti-seismic regulations.

\begin{figure}
 \begin{subfigure}{7.25cm}
    \includegraphics[width = 7.5cm]{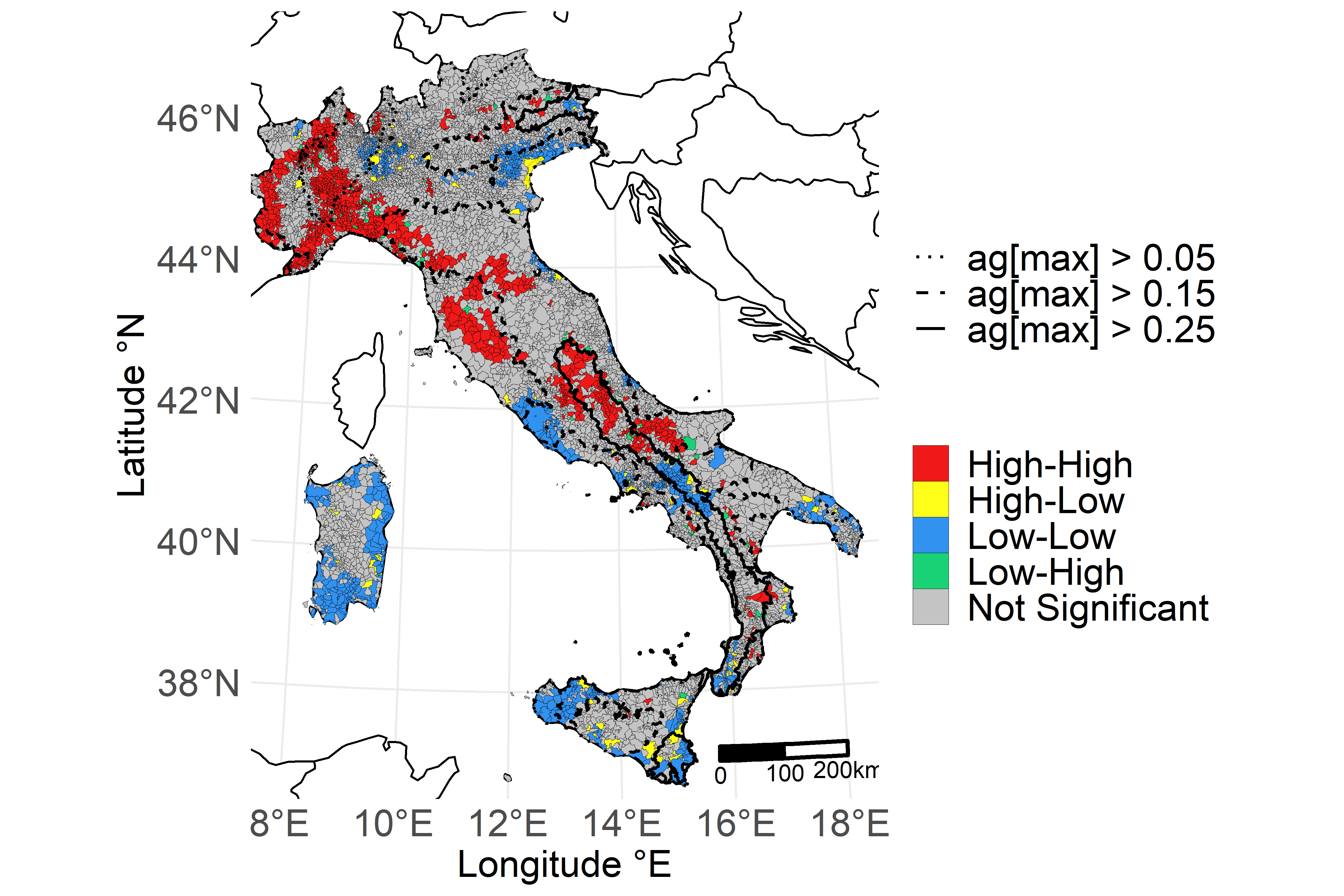}
\caption{ }
  \end{subfigure}
  \begin{subfigure}{7.25cm}
    \centering
    \includegraphics[width = 7.5cm]{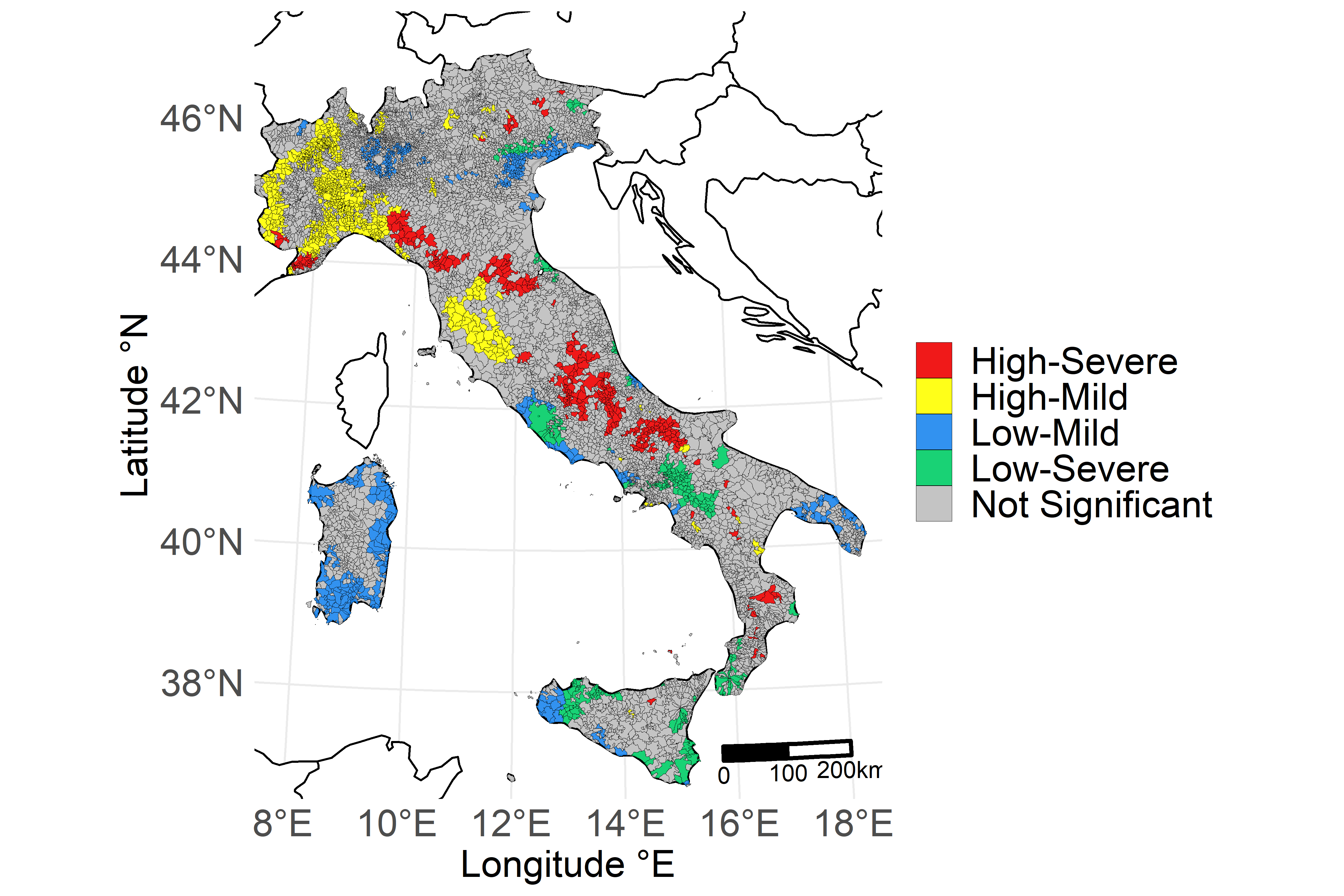}
\caption{ }
  \end{subfigure}
\caption{Building stock age. (a) Univariate LISA map of the \(PC1\) scores of the complete time-of-construction distributions; dashed lines represent the boundaries among classes of seismic hazard. (b) the High-High, Low-Low hot-spots of the univariate LISA map of the \(PC1\) scores of the complete time-of-construction distributions stratified by \(ag[max]\) class.}
\label{fig:LISA_bpc}
\end{figure}

\begin{figure}[h!]
    \centering\includegraphics[width = 7.5cm]{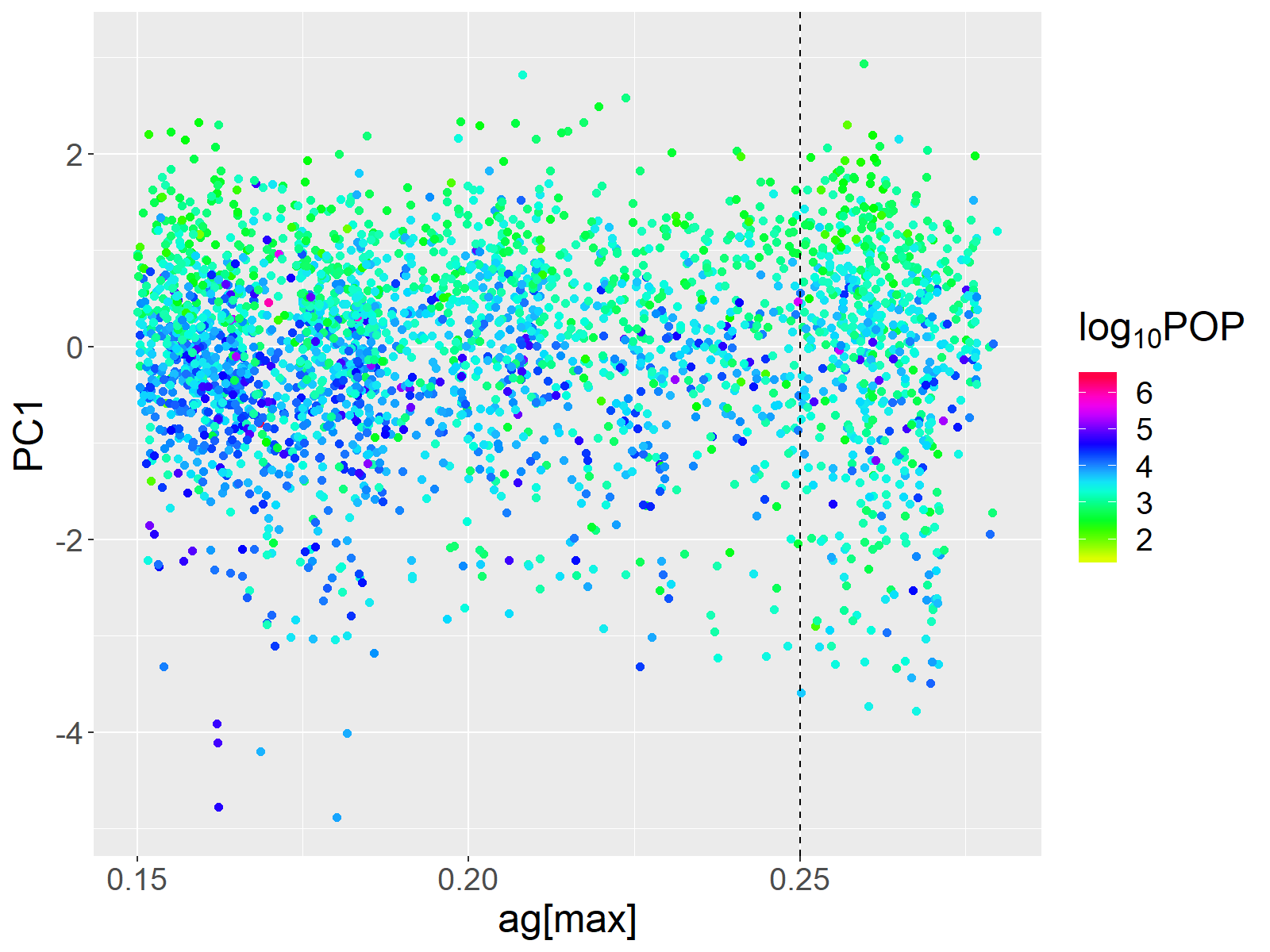}
    \caption{Building stock age: PC1 versus \(ag[max]\).}
  \label{fig:risk_Building_PC1}
\end{figure}

Coming back to the analysis of the complete time-of-construction distributions of the building stocks, data points of \(S^9,\) the global Moran`s I statistic measuring the spatial association of their  \(PC1\) scores is equal to \(0.506\).
To single out the country's hot spots with respect to the age of their building stock, we represent, in Figure \ref{fig:LISA_bpc}a, a univariate LISA map of the \(PC1\) scores  pictured in Figure \ref{fig:Building_PC1}a.
The map clearly points out the Low-Low clusters of more recent building stock: the large metropolitan attractors of demographic expansion around Milan and Rome, the North East, the coasts of Sardinia witnessing the development of its tourism industry, the already mentioned reconstructed Irpinia and Friuli regions, and notable spots in Sicily, Puglia and Calabria. The High-High hot spots identifying areas characterized by an ageing building stock run along the inner areas of the Appenines and also single out large portions of Piedmont, with the exception of the territory surrounding Turin, of Liguria and of Tuscany.
Figure \ref{fig:LISA_bpc}b segments these hot spots according to the two macro-classes of seismic hazard (severe/mild).
Whereas the green Low-Severe clusters indicate areas of the country where seismic hazard is severe but seismic risk is attenuated by a more recent building stock, concern is generated by the red High-Severe areas where a medium or high seismic hazard is associated with an ageing building stock. Once again these red areas belong to the inner regions of the country characterized by small communities, already suffering from a demographic decline and ageing of their residents. To elaborate on this point, consider Figure \ref{fig:risk_Building_PC1}. Each municipality appears as a point colored according to the \(\log_{10}\) of its resident population; on the abscissa the seismic hazard associated to the municipality, on the ordinate axis the \(PC1\) score. One notices immediately the overwhelming presence of communities with less than 10000 residents among those with extremely low values for the \(PC1\) score and therefore an ageing building stock. This evidence is even more compelling for those municipalities with a high seismic hazard (\(ag[max] > 0.25g\)).

\section{An aggregated analysis}\label{sect:aggregation}

Given the bounded resources constraint, policy makers need to prioritize their interventions against social and material vulnerability if they want to implement efficient precision policies. Indeed, ``the necessity to develop methods and indicators (to assess social vulnerability)... which can be used in policy and decision-making processes" has been stressed also in \citet{Birkman2006}. With this goal in mind, we try in this section to aggregate the results of the previous analyses with the aim of eliciting a conceivable global picture of social and material vulnerability in the face of seismic hazard for the Italian municipalities.

\begin{figure}
\begin{floatrow}
\ffigbox[10cm]{
  \includegraphics[width=10cm]{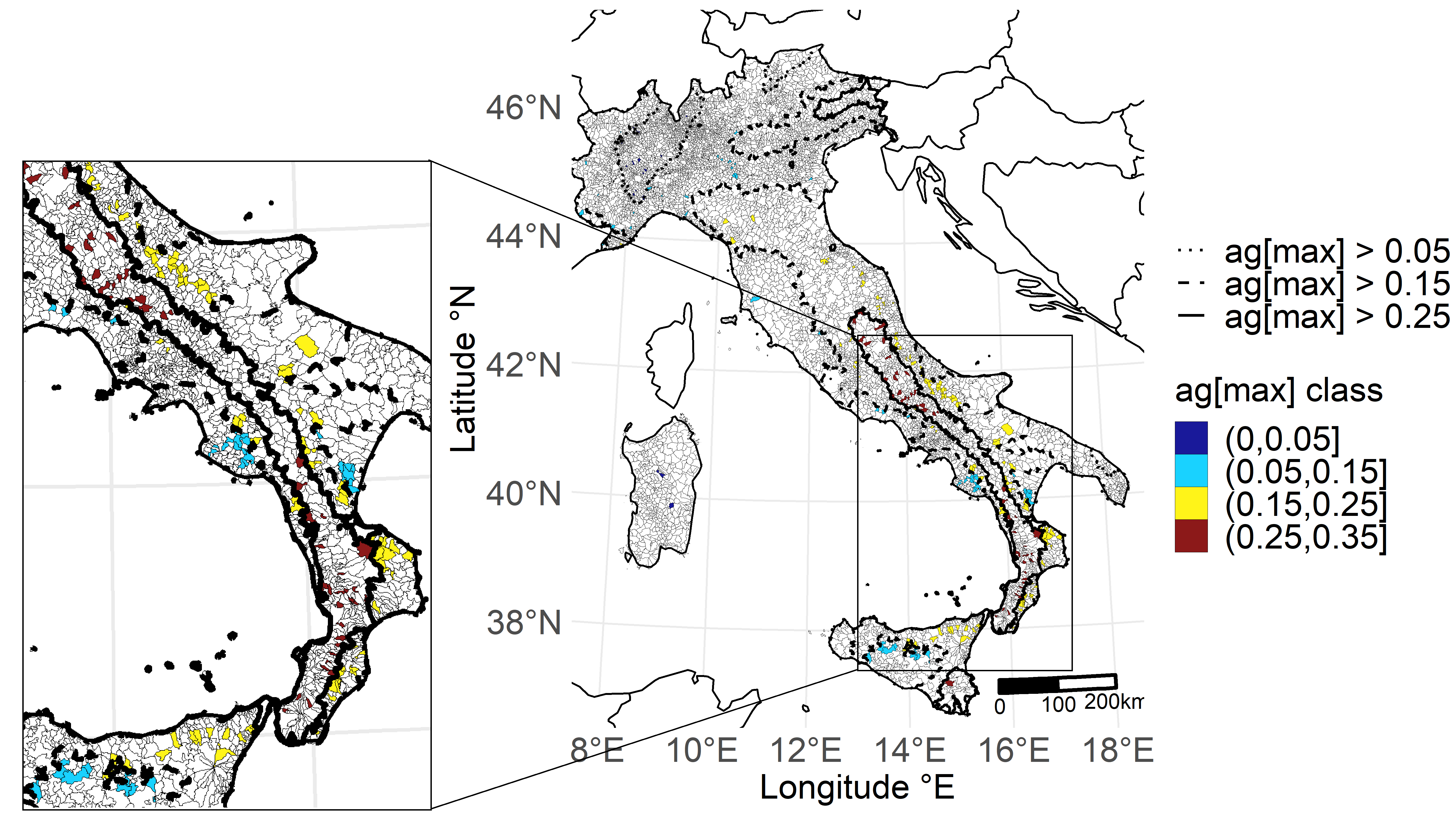}

}{%
  \caption[width = 5cm]{Map of Italy where each municipality selected according to criterion (8) has been colored according to \(ag[max]\) class.} \label{fig:quadrant}
}
\quad
\capbtabbox[4cm]{%
\begin{tabular}{ll}
\hline
\textbf{\(ag[max]\) class} & \textbf{Proportion} \\ \hline
(0,0.05{]}          & 0.07                \\ \hline
(0.05,0.15{]}       & 0.26                \\ \hline
(0.15,0.25{]}       & 0.46                \\ \hline
(0.25,0.35{]}        & 0.21                \\ \hline
&\\
&\\
&\\
&\\
\end{tabular}
}{%
  \caption{Proportion of municipalities selected according to criterion (8) for each class of $ag[max]$.\label{table:selected_mun_dist}}%
}
\end{floatrow}
\end{figure}

One obvious line of attack would be to single out those municipalities simultaneously troubled by a high \(IVSM\) value, demographic decline and an old building stock. To this effect, we set the threshold \(t_{IVSM} = 100.29\) equal to the third quartile of the \(IVSM\) distribution over the Italian municipalities, the threshold \(t_{g} = -4.64 \% \) equal to the first quartile of \(VAR\_PERC\)  and finally the threshold \(t_b = 1.18\) equal to the third quartile of the \(PC1\) score of the complete time-of-construction distributions of the building stock. Given the collinearity between \(VAR\_PERC\) and \(ETA\_Q3\) we decided to use only one of the two variables in the criterion for selecting municipalities; we settled on the former because it allows a better outlook on the future fragility of a community as it captures the demographic dynamics of its resident population.  We then highlight in Figure \ref{fig:quadrant} those municipalities with:

\begin{equation}\label{selection_criterion}
IVSM \geq t_{IVSM} \;\, \mbox{and}\;\, VAR\_PERC \leq t_g\;\, \mbox{and}\;\, PC1 \geq t_b.
\end{equation}

Out of 7953  municipalities, 247  pass the selection criterion \eqref{selection_criterion}. This supports the assumption of a very mild dependence for the three variables \(IVSM\),  \(VAR\_PERC\) and \(PC1\) since, in case of independence, we expect only \(125 \pm 23\) municipalities to pass \eqref{selection_criterion}.
The selected municipalities, highlighted in Figure \ref{fig:quadrant}, are colored according to their membership to one of the four classes of seismic hazard as defined in Section \ref{sect:seismic_landscape}, and identified by the contour lines on the plot. Their distribution in the four seismic classes is reported in Table \ref{table:selected_mun_dist}; one may notice that  for almost 70\% of them, the seismic hazard is medium or high.

A second line of attack aims at a global ranking of the Italian municipalities in line with the three indexes we have so far considered for capturing social and material vulnerability; for this, we will use the Copeland method illustrated in \citet{copeland}. First, we separately rank the municipalities consistently with \(IVSM\),    \(VAR\_PERC\) and the  \(PC1\) scores of their complete time-of-construction distributions of building stock. For ranking municipalities according to \(IVSM\) and \(PC1\), we proceed in ascending order, attributing rank 1  to the municipality with the smallest index, that is the best in terms of \(IVSM\) or \(PC1\) score, respectively. Ranking of municipalities according to \(VAR\_PERC\)  proceeds instead in descending order, rank 1 being attributed to the municipality having highest \(VAR\_PERC,\) that is the best in terms of demographic growth. Ties are sorted by population size, a smaller population size inducing a larger rank number.
In other words, in each of the three rankings, larger rank numbers are indicative of a higher vulnerability, as captured by the indexes analyzed in the previous sections of this paper. To obtain a final global ranking, we then consider all pairwise competitions between municipalities. When municipality \(i\) competes against municipality \(k\),
let \(r_{il}\) and \(r_{kl}\) be their respective rank numbers in the rankings  \(l=1,2,3,\) described above and set the result of the competition to be
\begin{equation}
    s_{i,k}=\left\{
                \begin{array}{ll}
                  1,  & \text{ if  } \#\{l \in \{1,2,3\}: r_{il}>r_{kl}\} \geq  2; \\
                 -1, & \text{ if  } \#\{l \in \{1,2,3\}: r_{il}>r_{kl}\}  \leq 1.
                \end{array}
                \right.
\end{equation}
Finally, for \(i=1,...,n,\) we compute the Copeland score of the municipality  $i$ as the sum over the results of all pairwise competitions,
 $$
 C_i = \sum_{k \neq i} s_{i, k}.
 $$

\begin{figure}
    \centering\includegraphics[width = 10cm]{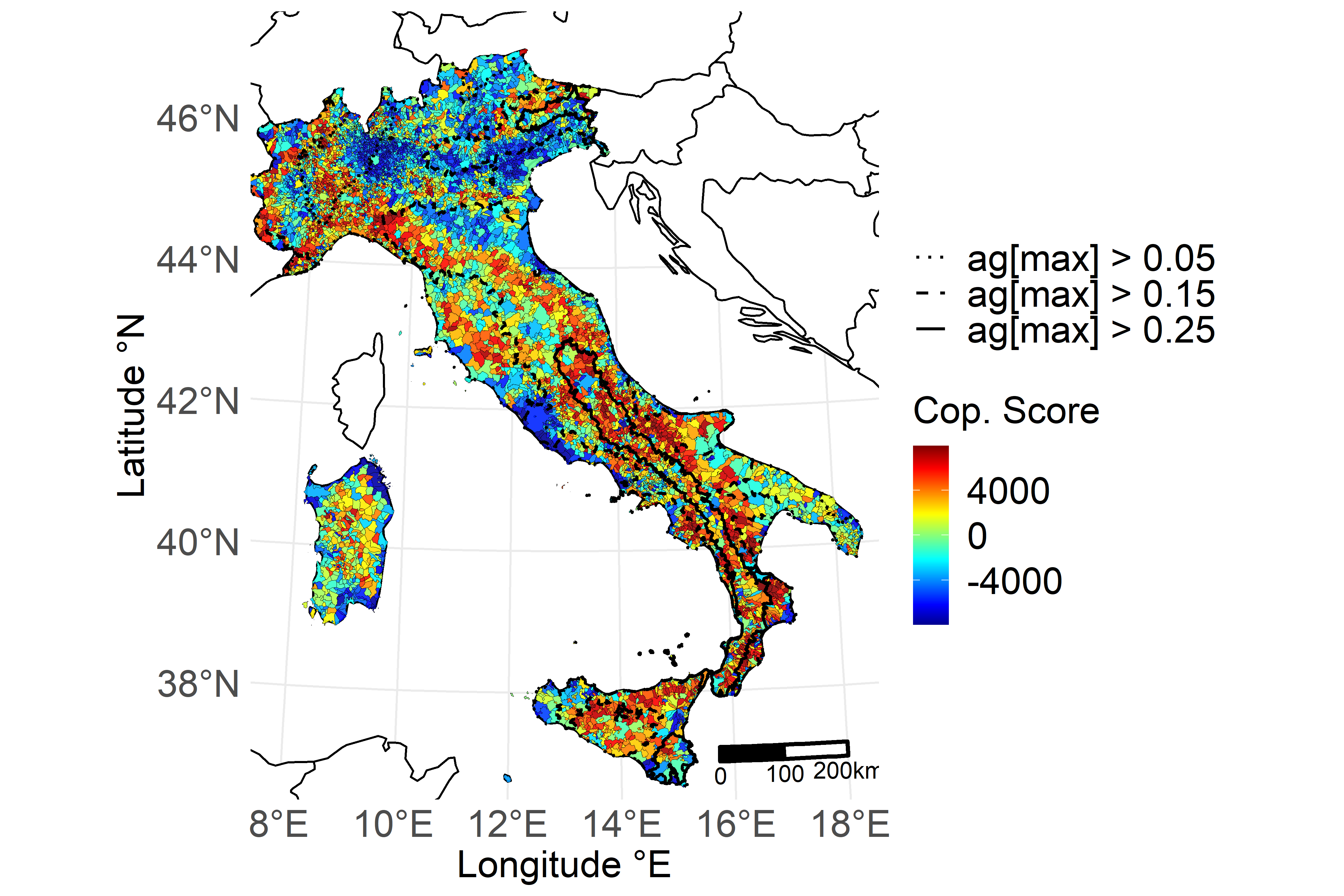}
  \caption{Map of Italy where each municipality has been colored according to the Copeland score.}
  \label{fig:Italy_copeland}
\end{figure}

Figure \ref{fig:Italy_copeland} shows the map of Italy where each municipality has been colored according to its Copeland score, hot colors for the units with a high score, i.e. worse off in our global ranking of social and material vulnerability, cold colors for those municipalities with a low Copeland score, better off in terms of social and material vulnerability. The contours defining the four classes of seismic hazard are also reported.  We can see that the ranking provided by the Copeland score highlights vulnerable areas in Piedmont, in Friuli and in the inner regions of Apennines and of the main islands, consistently with our previous analyses. 
In particular, the spatial arrangement of the high-priority areas evidenced by our ranking is also overall coherent with that of the candidate pilot zones selected by the Italian Government within the national program for the Inland Areas. \footnote{
Ministro per la Coesione Territoriale e il Mezzogiorno Claudio De Vincenti (2016), Relazione annuale sulla
Strategia nazionale per le aree interne, p. 15; available at \url{www.programmazioneeconomica.gov.it/2017/02/08/relazione-annuale-sulla-strategia-nazionale-per-le-aree-interne-4}
}. This provides further evidence that gentrification still appears to be a lateral phenomenon with respect to the overall vulnerability of these areas, despite their key cultural and historical value. 
 Our results are overall consistent with the classes of vulnerability identified by \citet{Frigerio16b}, although some discrepancy appears in the detailed comparison of the results (see Fig. 5 of \citet{Frigerio16b}). In fact, our results suggest a stronger contrast between inland and outer regions, and a higher vulnerability of the municipalities in the Piedmont region, possibly due to a non-negligible effect of the building stock information in our analyses. Nevertheless, similarly as for \citet{Frigerio16b}, our results suggest that the overall ranking of Italian municipalities based on social and material vulnerability well reflects the national topography, which in turn is strongly associated with the phenomenon of desertion of mountains areas in favor of more industrialized and urbanized ones.

\begin{figure}
    \centering\includegraphics[width = 7.5cm]{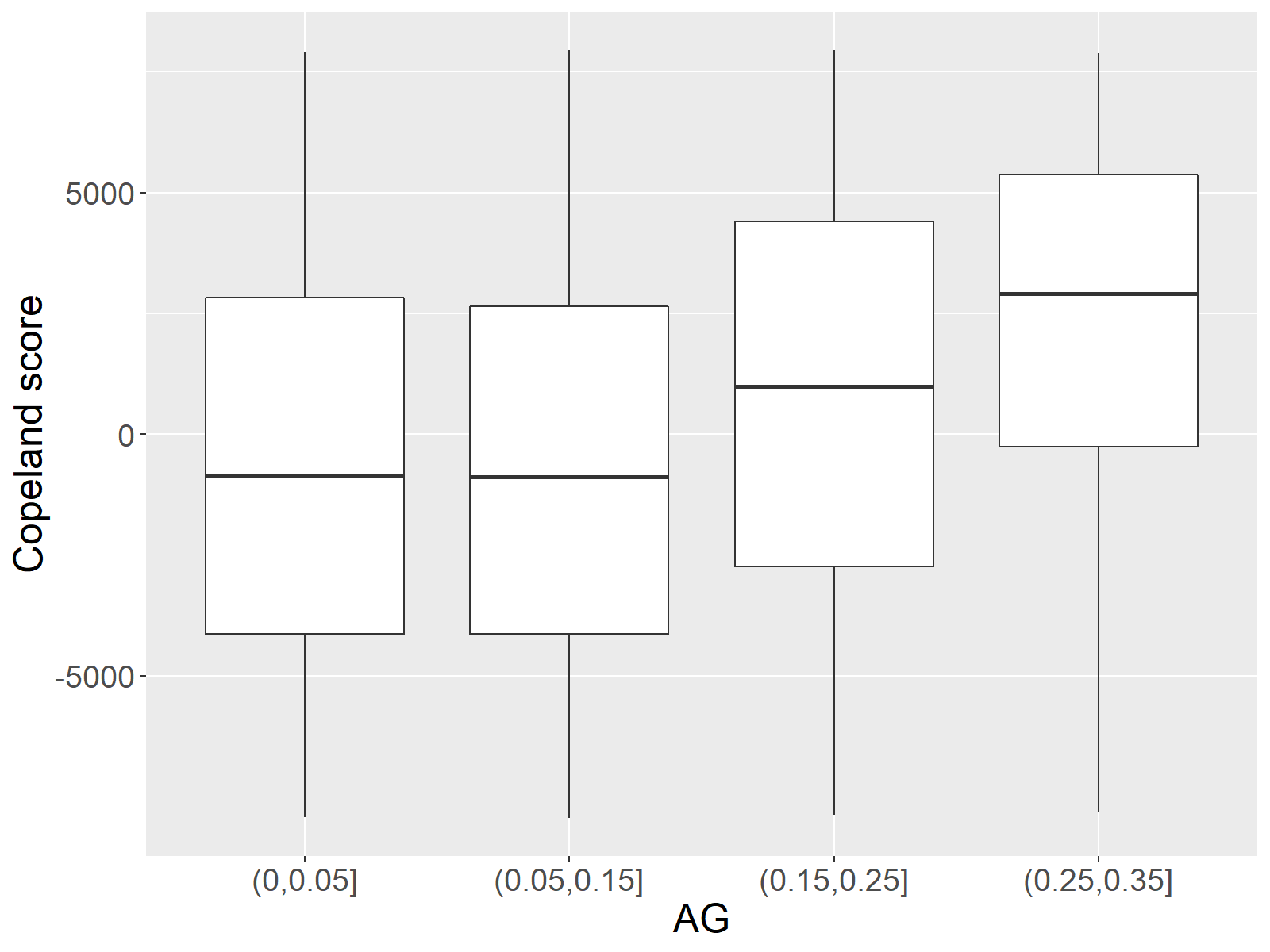}
  \caption{Conditional distributions of Copeland score, given \(ag[max].\)}
  \label{fig:cop_boxplt}
\end{figure}

One may also notice that the areas characterized by high Copeland scores are frequently concurrent with the areas of severe seismic hazard 
(see, e.g., the inner areas in Abruzzo and Calabria).
 To elaborate on this consider Figure \ref{fig:cop_boxplt} which depicts as boxplots the conditional distributions of the Copeland score, given the values of \(ag[max]\) defining the four seismic classes listed in Section \ref{sect:seismic_landscape}. There is a clear stochastic order among the four distributions, those associated to larger values of \(ag[max]\) dominating those corresponding to less hazardous areas. Therefore, for municipalities belonging to the medium or high seismic hazard classes, social and material vulnerability is acting as an amplifier of seismic hazard, contributing to the increase of risk instead of dimming it.

Ultimately, ranking through the Copeland's score allows to prioritize the public intervention on Italian municipalities, with the goal of reducing their vulnerability to seismic hazard. Each community is however different, and the multidimensional analysis carried out in the previous sections allows to discriminate among the components which differentially contribute  to the social and material vulnerability of Italian municipalities, profiling each municipality in terms of its dominant fragilities when exposed to a seismic event, and allowing decision makers to plan tailored precision policies to fight them.   This is in contrast to the actual policies designed to reduce seismic risk, which focus only on the hazard factor and cluster the municipalities in terms of the four hazard classes listed in Section \ref{sect:seismic_landscape}, promoting and incentivizing prevention against the risk by means of equal measures shared by all municipalities belonging to the same cluster.

\section{Further supporting analyses}
In the previous section, we claimed that $VAR\_PERC$  allowed a better outlook on the future fragility of a community because it captures the demographic dynamics of its resident population. We thus used this variable, instead of the correlated $ETA\_Q3$, when building a global ranking of the Italian municipalities in terms of social and material vulnerability.  Although MRCI provides, through $VAR\_PERC$, the variation of the number of residents between 2011 and 2018, the resident population of every municipality is constantly monitored by ISTAT. In fact, based on the data made public by ISTAT, for every Italian municipality we know the annual demographic growth between 1992 and 2012.

\begin{figure}[h]
  \begin{subfigure}{7.25cm}
    \centering\includegraphics[width=7.5cm]{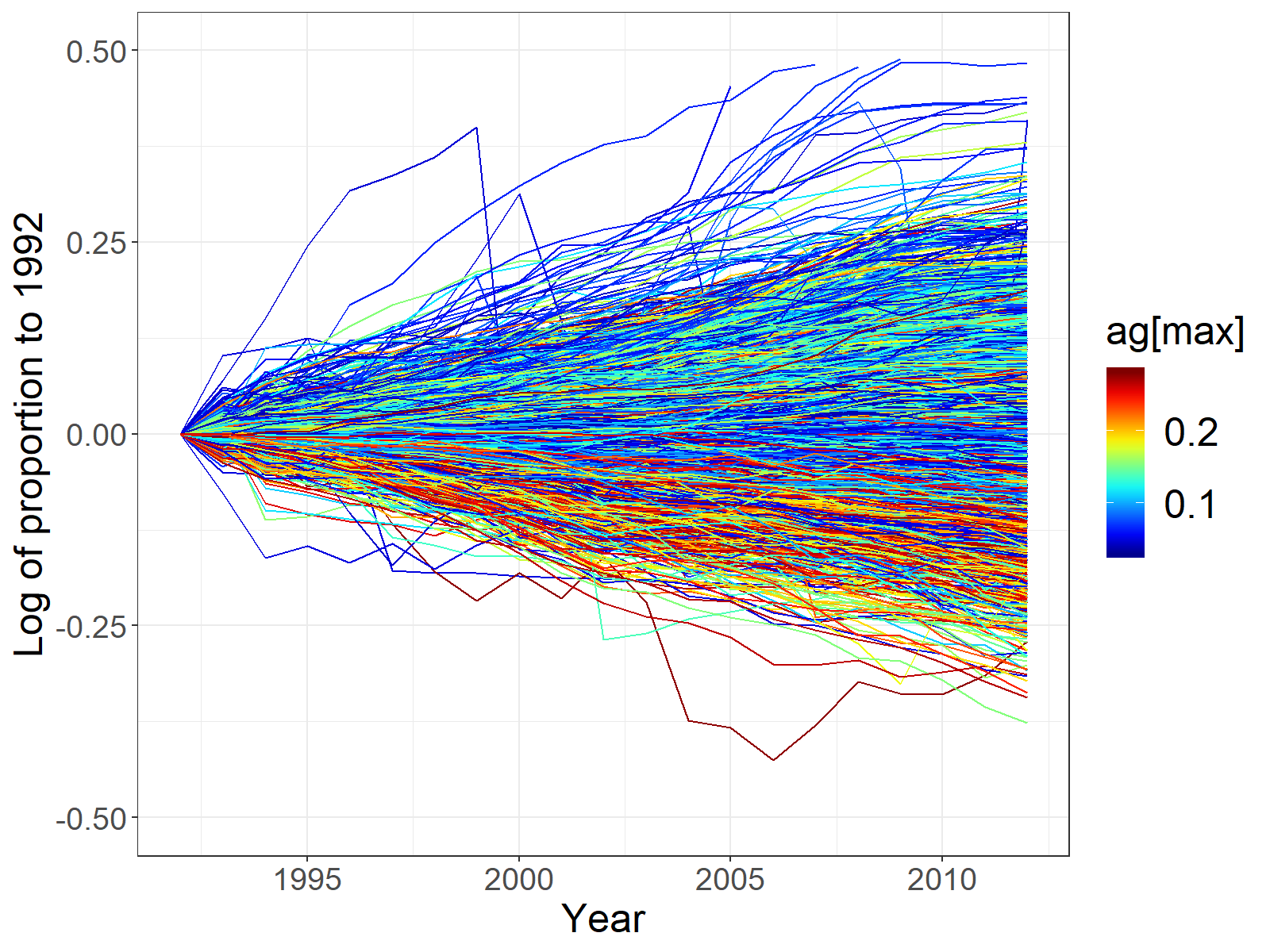}
    %\caption{}
  \end{subfigure}
  \begin{subfigure}{7.25cm}
    \centering\includegraphics[width=7.5cm]{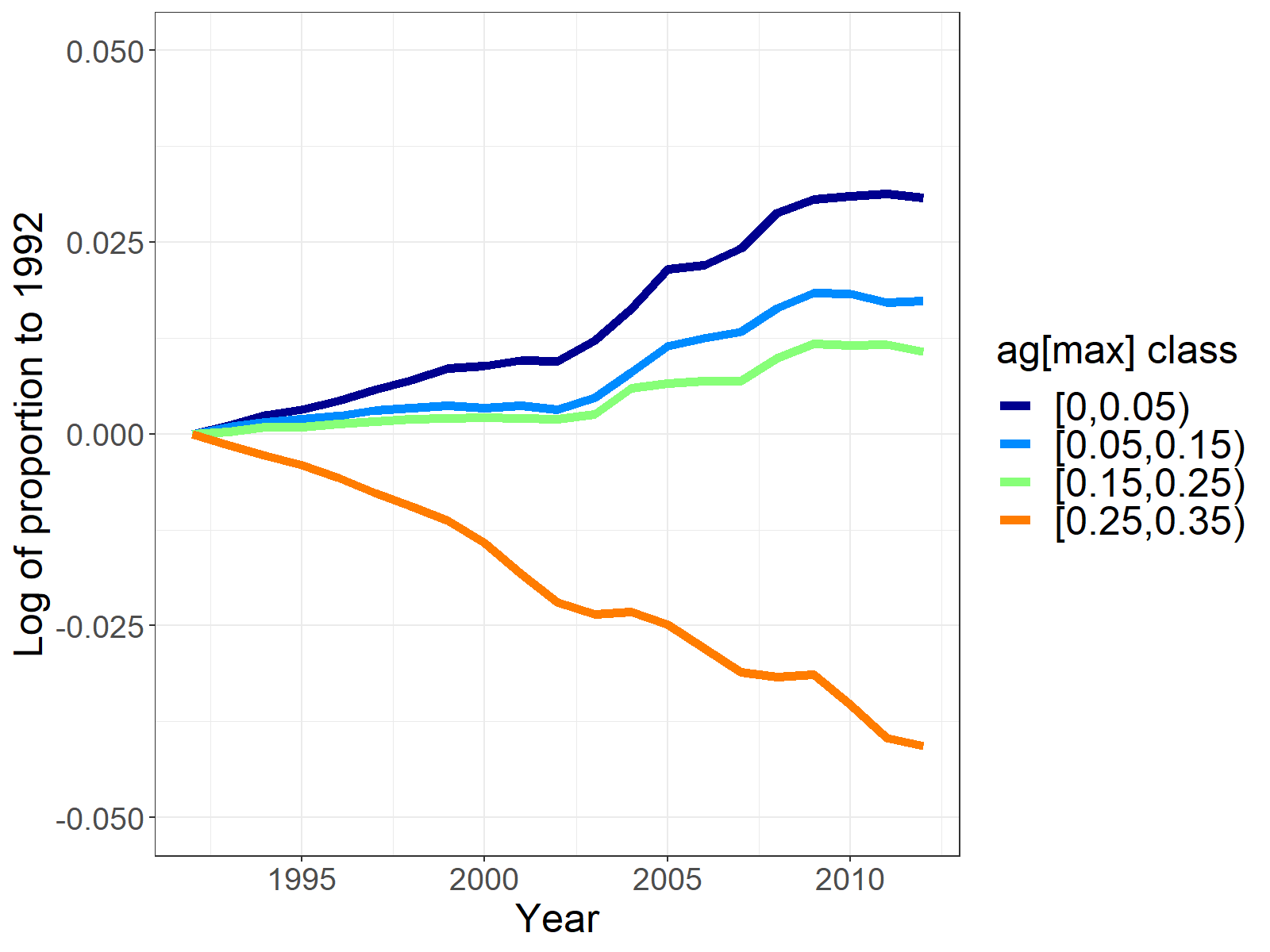}
    \caption{}
  \end{subfigure}
  \caption{Temporal population dynamics: in (a) each municipality is represented by the log growth curve, colored according to the \(ag[max]\). In (b) curves are year-wise medians of the log growth rates for the four \(ag[max]\) classes. Note that different scales are used for the y-axes in plots (a) and (b), to enhance visibility of the curves.}
  \label{fig:growth_rates}
\end{figure}

Figure \ref{fig:growth_rates}, on the left, shows for each Italian municipality the \(\log\) of the ratio between its population size in a given year and its population size in 1992, taken as reference year, that is the \(\log\) of its demographic proportional growth with respect to 1992. Each curve is colored according to the \(ag[max]\) associated to the municipality. One might notice that for many municipalities troubled by a medium or high seismic hazard the population is not increasing, when it is not steadily declining. Indeed this is confirmed by the graph on the right where, for the four seismic hazard classes, the medians of the \(\log\) proportional growths are represented for every year. 

\begin{figure}
  \begin{subfigure}{7.25cm}
    \centering\includegraphics[width=7.5cm]{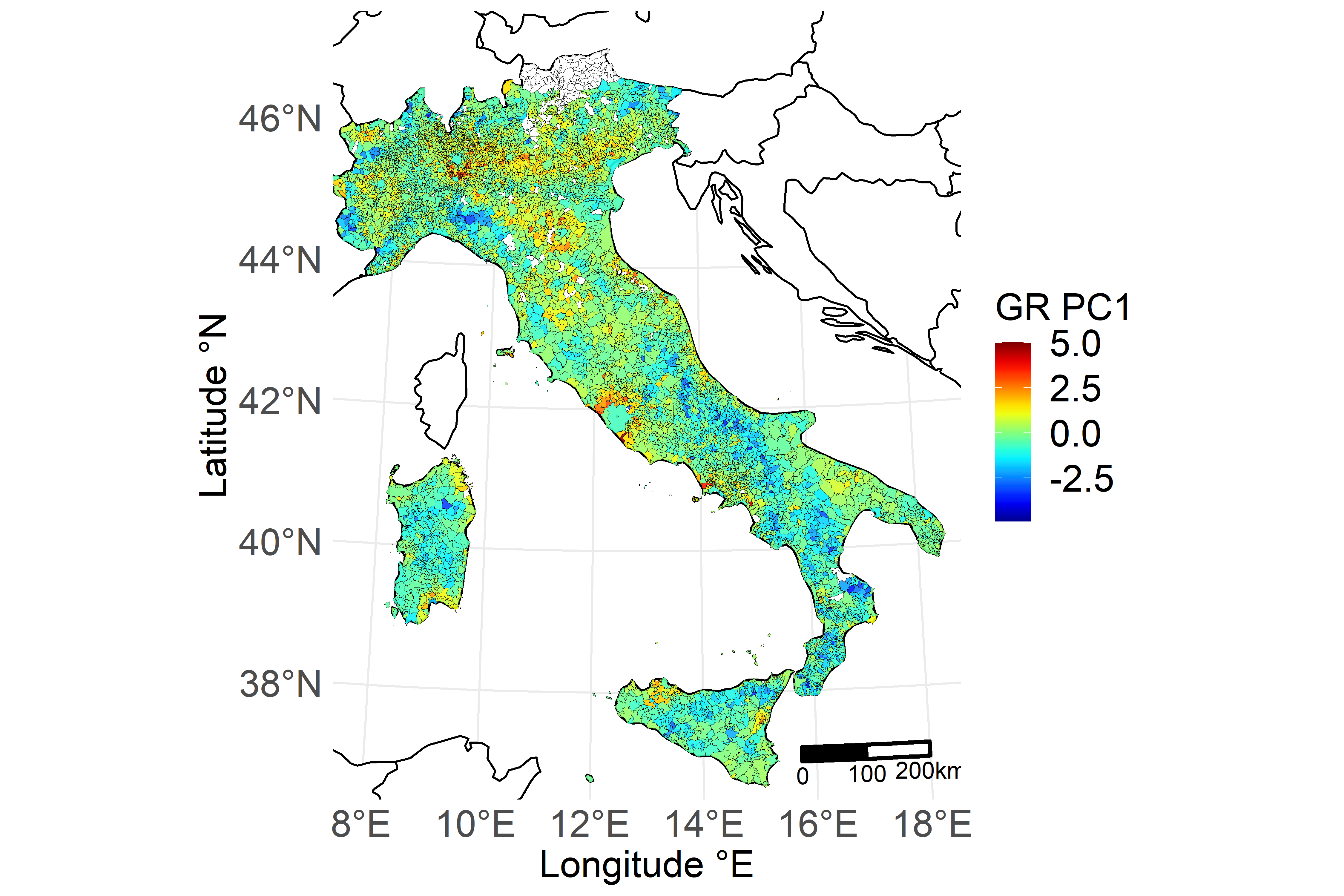}
    \caption{}
  \end{subfigure}
  \begin{subfigure}{7.25cm}
    \centering\includegraphics[width=6.7cm]{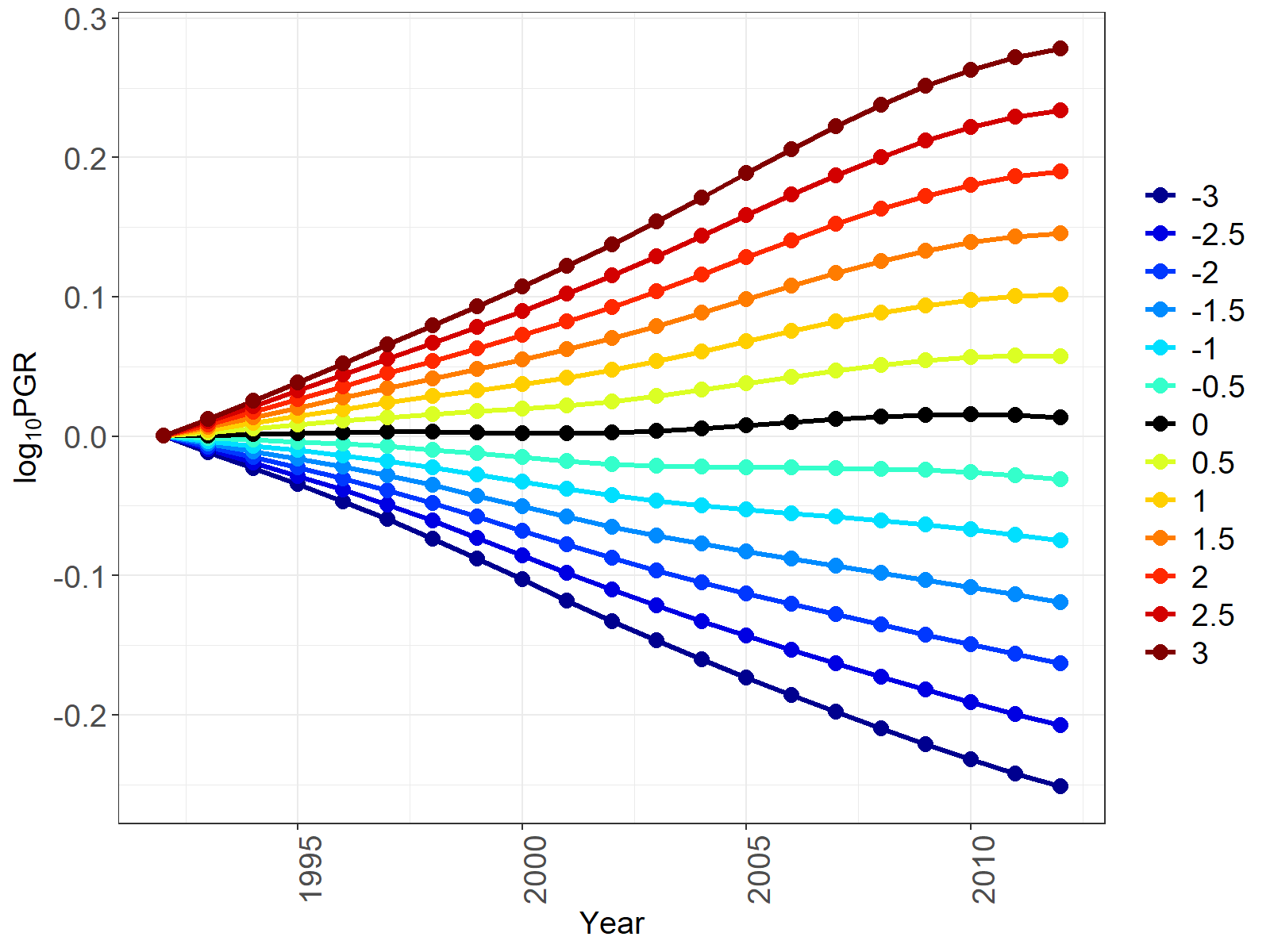}
    \caption{}
  \end{subfigure}
  \caption{Proportional growth curves: in (a) each municipality has been colored according to the the score of its \(\log\) proportional growth curve on the first Functional Principal Component. The score color identifies the corresponding \(\log\) proportional growth curve on the guide chart (b).}
  \label{fig:log_gr_fpca}
\end{figure}

After smoothing the \(\log\) of the demographic proportional growth curves by means of 11 cubic B-splines, we performed a Functional Principal Component Analysis \citep{FDA_book}. The first principal component captures 95.4$\%$ of the data variability, thus supporting their projection on a 1-dimensional linear subspace, where each curve is encapsulated in a real number, its score on the first principal component.
Figure \ref{fig:log_gr_fpca} shows the map of Italy, where each municipality has been colored according to these scores; on the right the guide chart for interpreting the values of the scores, cold colors being associated to declining \(\log\) proportional growth curves, hot colors otherwise. The black curve is the average \(\log\) proportional growth curve for the entire country. These scores are correlated (\(\rho = 0.94\)) to the variation index $VAR\_PERC$ reported in the MRCI and analyzed in the previous sections, thus supporting the claim that $VAR\_PERC$ is tracking a long term trend of the local demographic dynamics.

\begin{figure}
  \begin{subfigure}{7.25cm}
    \centering\includegraphics[height=4.5cm]{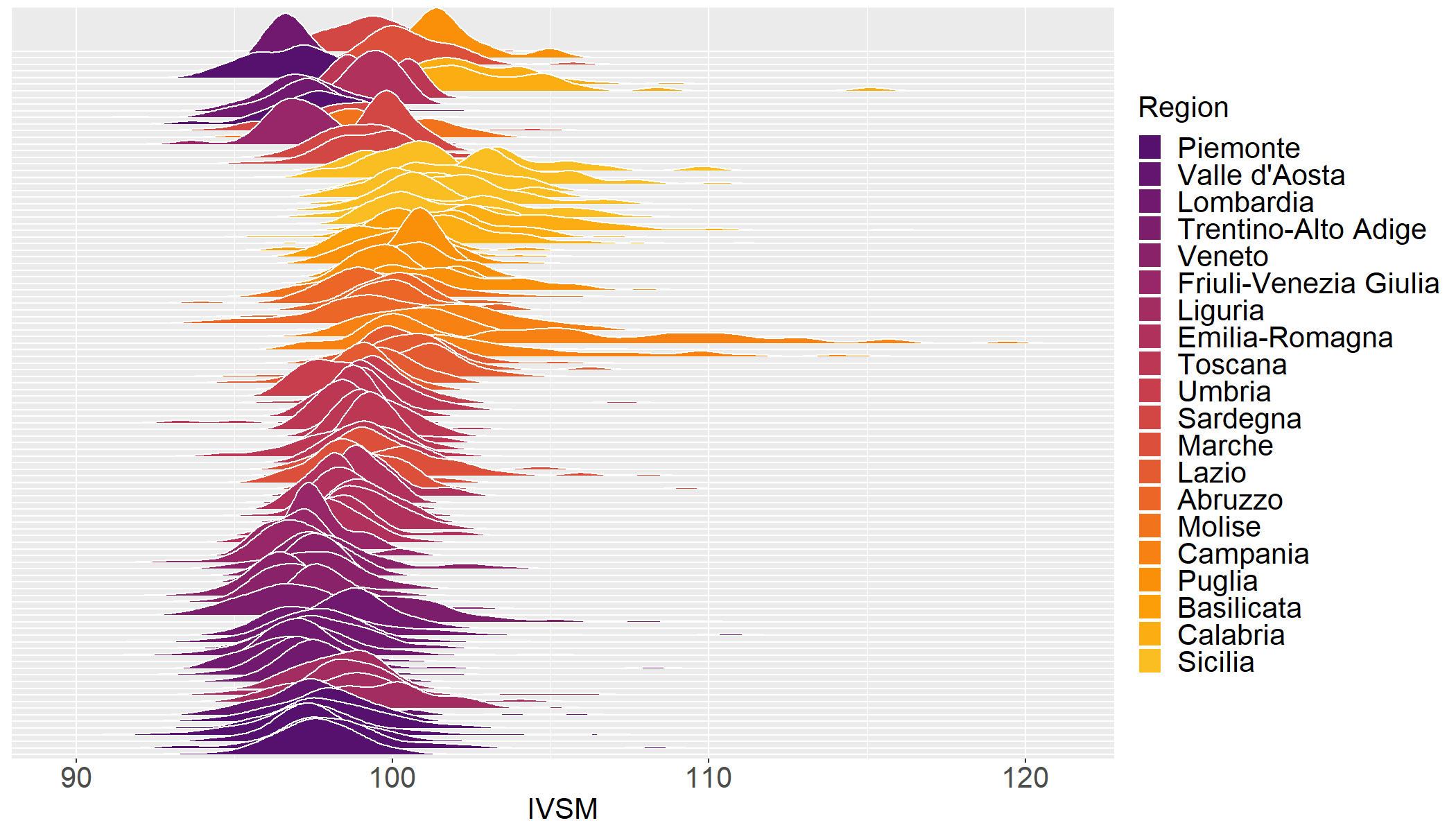}
    \caption{}
  \end{subfigure}
\begin{subfigure}{7.25cm}
  \centering\includegraphics[height=4.5cm]{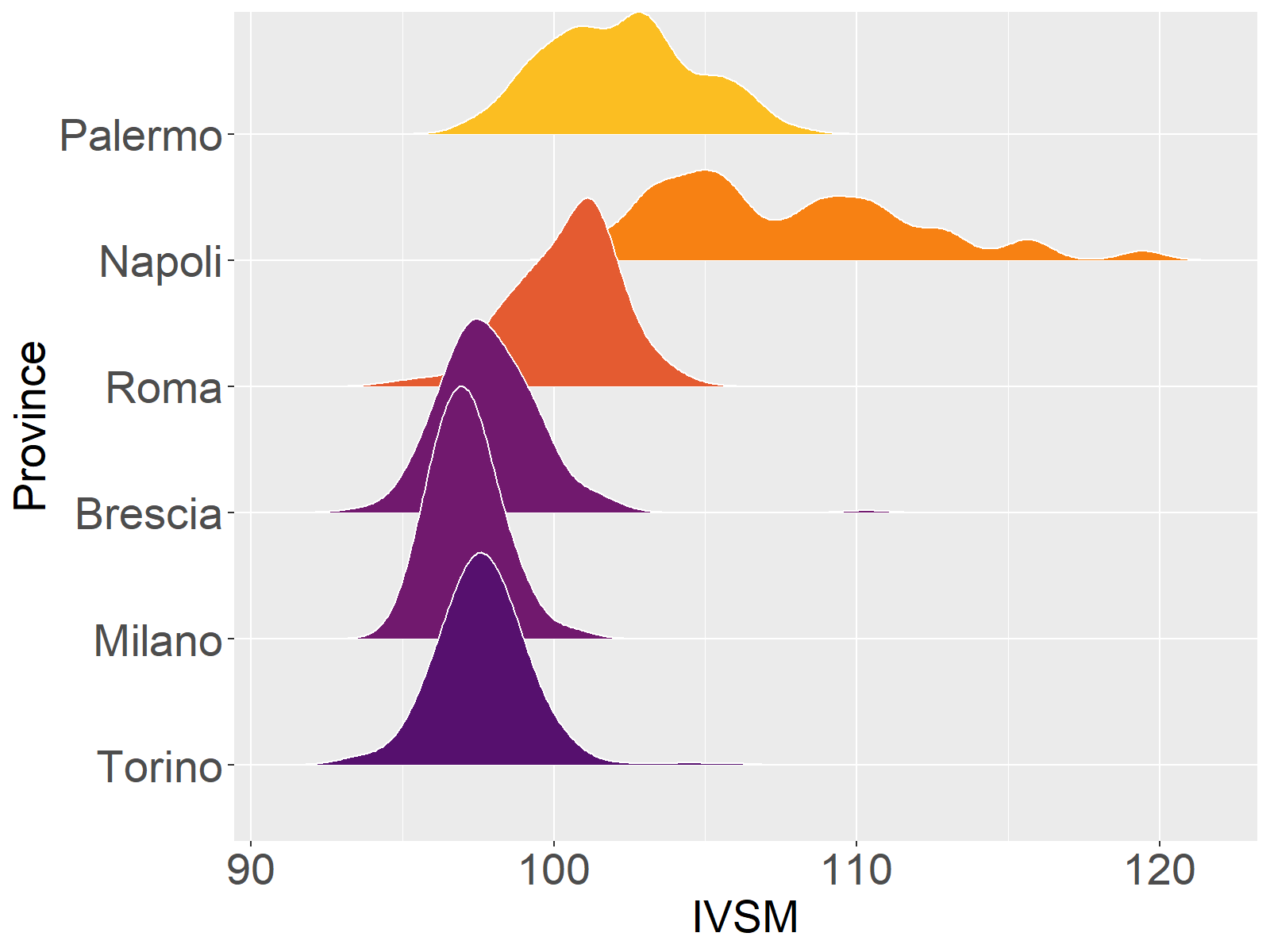}
\caption{}
  \end{subfigure}
  \caption{\(IVSM\) distributions for the Italian provinces. In (b) the six largest Italian provinces in terms of resident population, year 2018}
  \label{fig:IVSM_prov}
\end{figure}

A second point worth of discussion is the spatial variability of the index \(IVSM\), analyzed in Section \ref{sect:IVSM}. It was already noted that \(IVSM\) shows a marked trend moving from the North to the South of the peninsula. Although less obvious, the careful reader of Figure \ref{fig:IVSM_risk} will also notice that the variability of \(IVSM\) between municipalities sharing a similar latitude is also changing when moving form North to South. To better support this claim, we looked at the \(IVSM\) data at a different spatial scale, considering provinces instead of municipalities. In fact, for each Italian province we may compute the distribution density of the value of \(IVSM\) observed for its municipalities. For better visualization and treatment, these distributions are then smoothed via kernel smoothing (Gaussian kernel, bandwidth = 0.527); they are depicted in Figure \ref{fig:IVSM_prov}.

\begin{figure}[h!]
    \centering\includegraphics[width = 14cm]{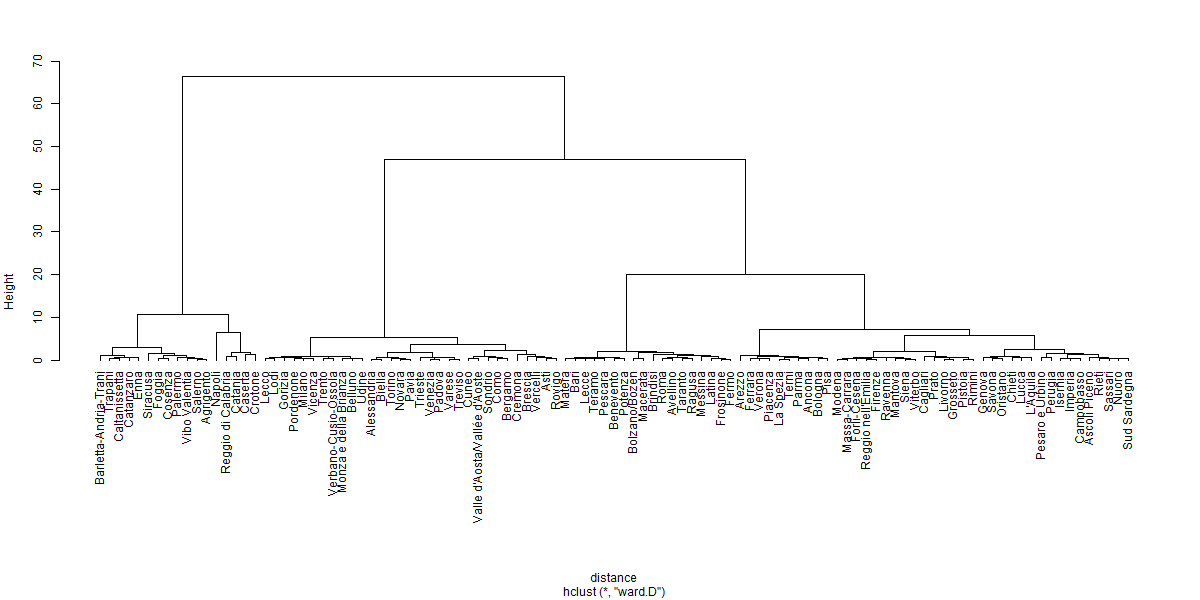}
  \caption{\(IVSM\): the dendrogram representing the clustering structure of the provinces}
  \label{fig:IVSM_dendr}
\end{figure}

\begin{figure}[h!]
\begin{subfigure}{7.25cm}
    \includegraphics[width = 7.5cm]{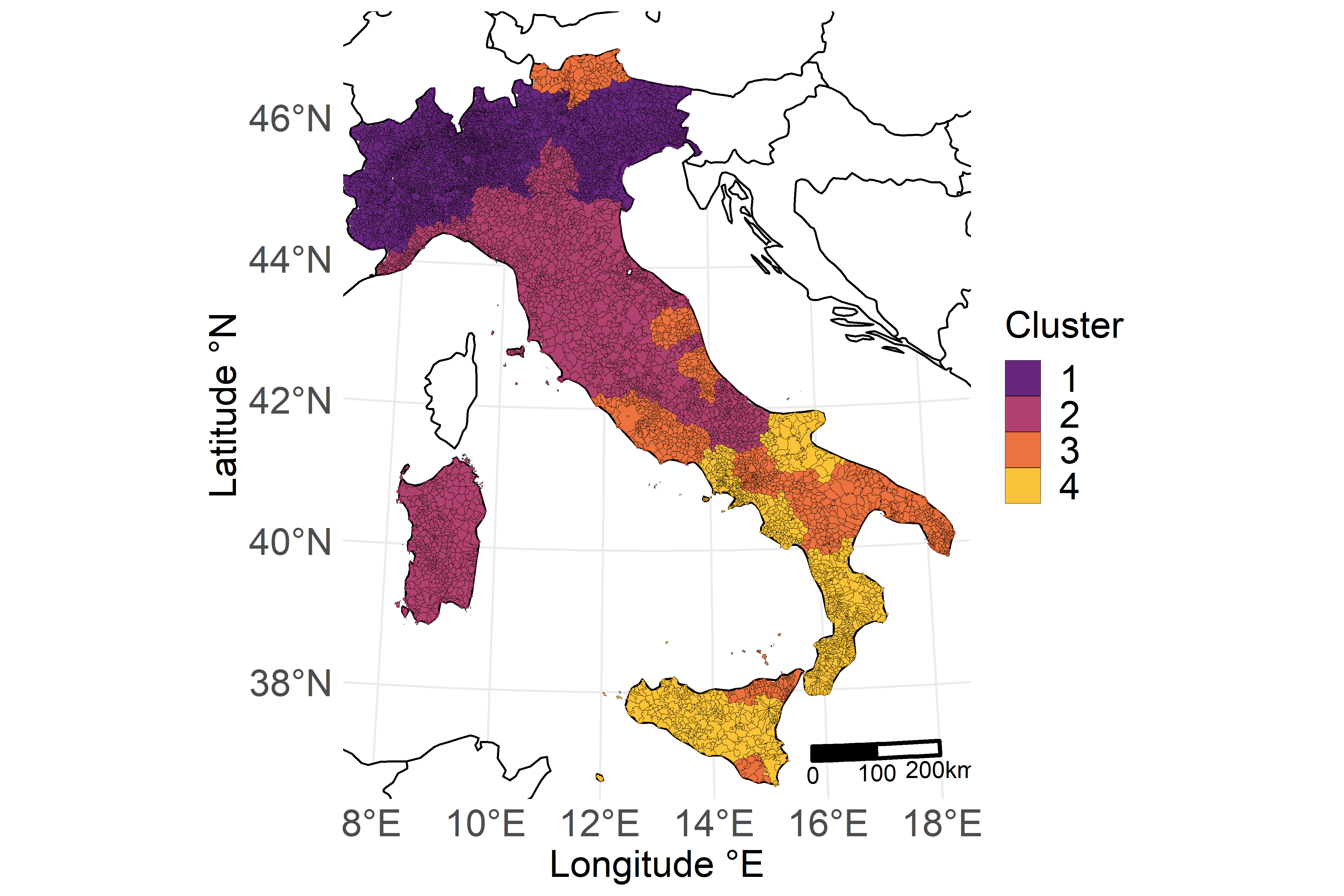}
     \caption{}
    \end{subfigure}
    \begin{subfigure}{7.25cm}
    \includegraphics[width = 7.5cm]{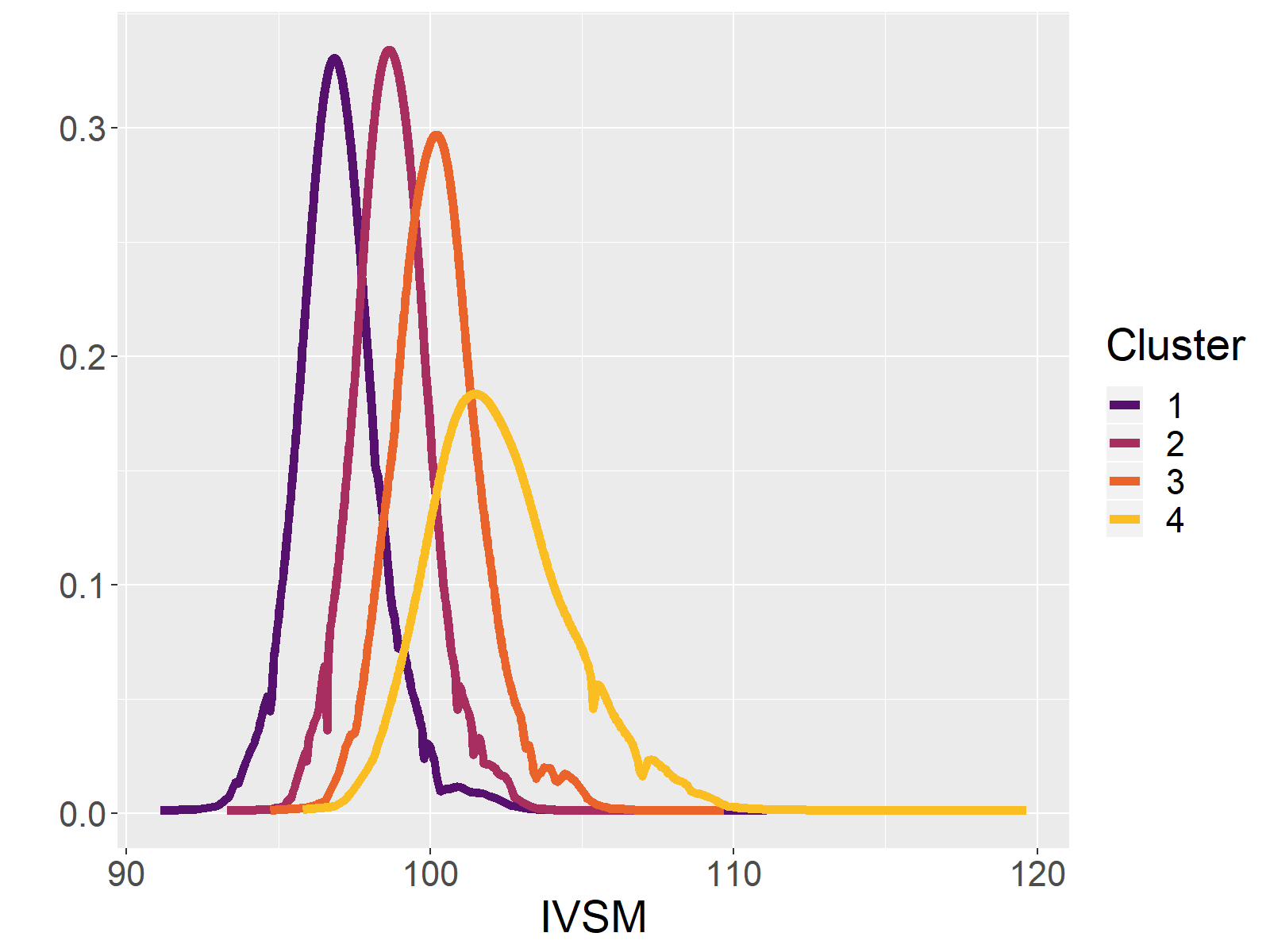}
    \caption{}
    \end{subfigure}
\caption{\(IVSM\): clustering of provinces using Wasserstein distance. In (b) barycenters of the four clusters}
\label{fig:IVSM_clt}
\end{figure}

A shift from dark purple to yellow corresponds to a shift from North to South, the colors being assigned on a regional basis. One may notice a strong spatial trend for both the location of the densities and their scale, when moving from North to South. This implies a large variability between provinces in terms of \(IVSM\), provinces in the South of Italy and Sicily being associated to larger values of mean \(IVSM\),  but also an increasing heterogeneity within the provinces, the provinces in the South being characterized by a larger variability of the \(IVSM\) of their municipalities. To make this trend even more manifest, we ultimately explore the classification of provinces based on the distribution of \(IVSM\) among their municipalities. For any pair of provinces \((P,Q),\) we compute the Wasserstein distance between the cumulative distributions \(F_P\) and \(F_Q\) of the \(IVSM\) for their municipalities:
 \[
d_W(P,Q) = \sqrt{\int_0^1 [F_P^{-1}(t) - F_Q^{-1}(t)]^2 dt}.
\]
We then run the Agglomerative Hierarchical Clustering Algorithm with Ward linkage to obtain the dendrogram plotted in Figure \ref{fig:IVSM_dendr}. Cutting the dendrogram to obtain four clusters, we get the clusters of provinces represented in Figure \ref{fig:IVSM_clt}, on the left, which seems to be in strong connection with the representation of \(IVSM\) offered by Figure \ref{fig:IVSM_risk}. On the right of Figure \ref{fig:IVSM_clt}, the barycentric distributions of the three clusters are reported; they clearly support the claim of a North-South regional trend for the \(IVSM\) provincial distributions, both in terms of location and of scale. The claim is also quantitatively supported by Table \ref{table:mean_var} where mean, standard deviations, and quartiles of these barycentric distributions are reported.

\begin{table}[h!]
\begin{tabular}{clllll}
\hline
\multicolumn{1}{l}{\textbf{Cluster}} & \textbf{Mean} & \textbf{sd} & \textbf{Q1} & \textbf{Q2} & \textbf{Q3} \\ \hline
1                                      & 97.09         & 1.74        & 96.13       & 96.93       & 97.80       \\ \hline
2                                      & 98.80         & 1.44        & 97.92       & 98.72       & 99.54       \\ \hline
3                                      & 100.29        & 1.58        & 99.29       & 100.20      & 101.13      \\ \hline
4                                      & 102.43        & 2.67        & 100.68      & 102.08      & 103.76      \\ \hline
\end{tabular}
  \caption{Summary indices of \(IVSM\) within the clusters of provinces.\label{table:mean_var}}%
\end{table}

\section{Discussion and Conclusion}

Social vulnerability is an abstract construct having multiple referents, none of which is fully comprehensive. By analyzing the data stored in the MRCI, we tried to capture a concept of social and material vulnerability which is functional to the design of precision policies for the prevention and reduction of the seismic risk of Italian municipalities. Encompassing \(IVSM,\)  we also considered as referents of social and material vulnerability the demographic dynamics of the communities and the age of the building stock where the community resides, grounding our argument on the belief that an aging and demographically declining community, living in buildings requiring significant improvements to be brought to safety, is more fragile and vulnerable when confronted with a seismic catastrophic event. Fusing measurements of different  factors associated to the latent social and material vulnerability, we formulated a tentative ranking of Italian municipalities apt to prioritize the public intervention for prevention and reduction of their seismic risk by means of precision policies rooted in their different fragility profiles, elicited by our multidimensional analysis.

Such approach supports a strategic re-design of policies aimed at facing seismic hazard, increasing their impact, efficiency and effectiveness. These policies must act on the vulnerability of buildings, as they cannot influence the probability or magnitude of earthquakes. In Italy, most buildings are privately owned; hence, public policies aim at stimulating owners to improve the safety of their houses. Usually, policies operate through tax deductions, available to all citizens living in areas with moderate to high seismic hazard. In 2017, for example, people were allowed a tax deduction of as much as 85\% of their investment. In spite of such a large incentive, however, the actual private investment for reducing the vulnerability of buildings was limited, for different reasons in different places. On the one hand, people living in places with moderate seismic hazard did not invest, as they considered even an 85\% funded investment not justified by the risk of an earthquake. On the other hand, most people living in socially vulnerable places do not take advantage of tax deductions or they cannot anticipate the required funds, due to low income, age, economic and welfare vulnerability. This problem is very critical  in places that are both socially vulnerable and with a high natural risk. Integrating the analysis of natural risk and social and material vulnerability, we are able to deal with the problem, moving from a “generic approach” to a “precision policy”, that modulates the amount and the form of the incentives according to the specific conditions of a single place. Focusing on a few critical places, it is indeed possible to increase the public support to private owners and anticipate the flows of money, with a limited impact on the National debt. For example, in Italy even a 100\% coverage of private investments by the Government, when limited to critical places, would increase the total amount of required funds by less than 1.2\% \mycitep{Report_2017}.

More in general, our work provides an analytical strategy for a global multi-aspect analysis, in which multiple viewpoints, possibly described by data of different nature (scalar, compositional, functional) can be analyzed jointly, enabling one to provide an overall ranking of the communities being studied, to eventually support decision makers. In this sense, the methodological approach here proposed is fully general, and could be easily complemented and enriched with additional sources of information on the various aspects characterizing seismic risk -- and possibly only partially covered within the MRCI. Among these, we mention the physical vulnerability of the buildings where communities reside, which could be explored and included in the analysis by considering the additional information regarding, e.g., the type of construction, the quality and state of conservation of the dwellings and the urban form (see, e.g., \citet{dolce2003earthquake,dolce2006vulnerability}). Here, a possible critical point could be the national coverage of the data, since part of this information may be available at a local scale only, as consequence of field campaign run at a county level (e.g., 
\citet{
d2002integrated,dolce2003earthquake,vicente2011seismic}). Another relevant aspect which may be worth considering is the daily dynamics of the population. As mentioned in Section \ref{sect:demographics}, data on the daily dynamics of the populations are typically indirectly collected, based on mobile phone data or mobility data, implying a number of issues, related with, e.g., uncertainty and selection bias. For these reasons, and following the rationale leading the activities of the Casa Italia task force, in this work we opted to base our analyses on \emph{certified} sources, meaning data collected by reliable institutions, with a national coverage (e.g., census data), as included in the public repository MRCI. We however foresee that accounting for these peculiar aspects of the population mobility -- as well as further aspects of the society being studied -- will be possible in a near future, eased by the large amounts of digital fingerprints left by the population in the daily activities. Indicators derived from these possible new sources of information could be naturally included in the strategy of analysis proposed in our work, as our approach is precisely aimed to allow for a multi-source, multi-aspect analysis.
\bibliography{CI_social_vulnerability}

\end{document}